\documentclass[12pt,english]{report}
\usepackage{amssymb, amsmath, amsthm}
\usepackage[hidelinks]{hyperref}
\usepackage{float}
\usepackage{xypic}
\usepackage[numbers]{natbib}
\usepackage{graphicx} %

\usepackage[left=1.7in, top=1.3in, right=1.3in, bottom=1.2in]{geometry}

\usepackage{setspace}
\usepackage{multirow}
\usepackage{threeparttable}

\theoremstyle{definition}

\theoremstyle{remark}

\numberwithin{equation}{chapter}
\numberwithin{figure}{chapter}

\usepackage[T1]{fontenc}
\usepackage[latin9]{inputenc}
\usepackage{geometry}
\usepackage{color}
\usepackage{amssymb,amsmath}
\usepackage{graphicx}
\usepackage{setspace}
\PassOptionsToPackage{normalem}{ulem}
\usepackage{ulem}
\usepackage{secdot}

\usepackage{tikz}
\usetikzlibrary{positioning}
\usetikzlibrary{arrows}
\usetikzlibrary{shapes.arrows}
\usetikzlibrary{matrix}
\usepackage{cellspace}

\makeatletter

\@ifundefined{showcaptionsetup}{}{%
 \PassOptionsToPackage{caption=false}{subfig}}
\usepackage{subfig}
\makeatother

\usepackage{babel}
    \makeatletter
    \def\thebibliography#1{\chapter*{References\@mkboth
      {REFERENCES}{REFERENCES}}\list
      {[\arabic{enumi}]}{\settowidth\labelwidth{[#1]}\leftmargin\labelwidth
	\advance\leftmargin\labelsep
	\usecounter{enumi}}
	\def\newblock{\hskip .11em plus .33em minus .07em}
	\sloppy\clubpenalty4000\widowpenalty4000
	\sfcode`\.=1000\relax}
    \makeatother
 
\newcommand{\myfigure}[3]{ \begin{figure}[tbp]
		\begin{center}
			#1
		\end{center}
		\caption{#2} \label{#3} \vskip -0.14in \end{figure} }

\newcommand{\mytable}[3]{ \begin{table}[tbp]
		\begin{center}
			\caption{#2}
			#1
			\label{#3} \vskip -0.01in \end{center} \end{table} }

\begin{document}

\pagenumbering{roman}
{

{\singlespacing

\newpage
\thispagestyle{empty}
\begin{center}
{ 
UNIVERSITY OF OKLAHOMA
\par
\vspace{0.16in}
GRADUATE COLLEGE
\par
\vspace{1.2in}
OPTIMAL FLOW ANALYSIS, PREDICTION AND APPLICATIONS\par
\vspace{0.17in}
\par
\vspace{1.2in}
A THESIS
\par
\vspace{0.17in}
SUBMITTED TO THE GRADUATE FACULTY
\par
\vspace{0.17in}
in partial fulfillment of the requirements for the
\par
\vspace{0.17in}
Degree of
\par
\vspace{0.17in}
MASTER OF SCIENCE
\par
\vfill
By
\par
\begin{singlespace}
	WEILI ZHANG
	\par
	Norman, Oklahoma
	\par
	2015
\end{singlespace}

}
\end{center}

\clearpage
\setcounter{page}{1}
\newpage
\thispagestyle{empty}
\begin{center}
{
OPTIMAL FLOW ANALYSIS, PREDICTION AND APPLICATIONS \par
\par

\vspace{0.9in}
A THESIS APPROVED FOR THE
\par
DEPARTMENT OF ENGINEERING
\par
\vspace{3in}
BY
\par
\vfill
\begin{flushright}
\begin{tabular}{cr}
\hline
\  \  \  \  \  \  \  & Dr. Charles D. Nicholson, Chair\\
 \\ &  \\
  \\ &  \\
\hline
  & Dr. Kash A. Barker \\
\\ & \\
 \\ &  \\
\hline
 & Dr. Suleyman Karabuk \\
\\   & \\
\end{tabular}
\end{flushright}
}
\end{center}

\newpage
\thispagestyle{empty}
\   \
\par
\vfill
\begin{center}
\copyright\ Copyright by WEILI ZHANG 2015

All rights reserved.
\end{center}
}

\newpage
\thispagestyle{empty}

\vspace*{\fill}
\begin{center}
\textit{This thesis is dedicated to my wife for her endless love, support and understanding.} 
\end{center}
\vspace*{\fill}
\newpage
\chapter*{Acknowledgments}

First and foremost, I would like to express my deepest gratitude to my advisor, Dr. Charles D. Nicholson, for his excellent guidance, patience, encouragement and providing me with an excellent laboratory for doing research. It has been a privilege for me to work with him. He has taught me all the necessary skills to be a good scientific researcher. I appreciate all his contributions of time and ideas that made my research journey productive and fascinating. Not only do his recommendations contribute to my research over the course of my studies, but his recommendations have also been a source of inspiration in many other aspects of my life. In addition, I would like to thank him due to coming to the gym and teaching me how to work out correctly.   

I gratefully acknowledge the funding sources provided by Dr.Nicholson and Dr. Naiyu Wang that made my graduate studies possible. Special thanks goes to Dr. Naiyu Wang who kindly hired me as her research assistant and funded me during my graduate studies. Her support and companionship were endless. My research would not have been possible without her kind support.  

I would also acknowledge Dr. Kash A. Barker who has always been supportive during the course of my studies. I am grateful to him for his enthusiasm for providing a friendly and enjoyable atmosphere in the School of Industrial and Systems Engineering. I also like to thank Dr. Suleyman Karabuk for accepting to be in my committee.    

Finally, I would like to appreciate all the faculties and staff for building the new Master degree of Data Science and Analytics, which is a fantastic and interesting program. It is my honor to be the first graduate student in this program. My time at the University of Oklahoma was made enjoyable in large part due to the many friends and groups that became a part of my life. My time at the University of Oklahoma was also enriched by getting along with absolutely nice faculty members, and amazing graduate and undergraduate students.      

\begin{spacing}{0.1}
\tableofcontents{}
\listoftables{}
\listoffigures{}
\end{spacing}

\chapter*{Abstract} 




This thesis employs statistical learning technique to analyze, predict and solve the fixed charge network flow (FCNF) problem, which is common encountered in many real-world network problems. The cost structure for flows in the  FCNF involves both fixed and variable costs. The FCNF problem is modeled mixed binary linear programs and can be solved with standard commercial solvers, which use branch and bound algorithm. This problem is important for its widely applications and solving challenges. There does not exist a efficient algorithm to solve this problem optimally due to lacking tight bounds. 

To the best of our knowledge, this is the first work that employs statistical learning technique to analyze the optimal flow of the FCNF problem. Most algorithms developed to solve the FCNF problem are based on the cost structure, relaxation, etc. We start from the network characteristics and explore the relationship between properties of nodes, arcs and networks and the optimal flow. This is a bi-direction approach and the findings can be used to locate the features that affect the optimal flow most significantly, predict the optimal arcs and provide information to solve the FCNF problem.

In particular, we define 33 features based on the network characteristics, from which using step wise regression, we identify 26 statistical significant predictors for logistic regression to predict which arcs will have positive flow in the optimal solutions. The predictive model achieves $88\%$ accuracy and the area under receiver operating characteristic curve is $0.95$. Two applications are investigated. Firstly, the predictive results can be used directly as component critical index. The failure of arcs with higher critical index result in more cost increase over the entire network. Specifically, in the $100$ instances, total cost has an average of $9.8\%$ increase while blocking highest two critical arcs. Secondly, we develop a regression-based relaxation (RBR) solution approach to the FCNF problem, in which the variable costs are replaced by a function of predictive probability. The rigorous experiments demonstrate the efficacy of the RBR by outperforming linear programming and state-of-the-art standard exact technique.

{\singlespacing
}
\newpage

} 


\pagenumbering{arabic}

\doublespacing
\chapter{Introduction}
\label{Introduction}

\section{Fixed Charge Network Flow Problem}
\label{Fixed Charge Network Flow Problem}

The fixed charge network flow problem (FCNF) was first developed by \citet{hirsch1954fixed}, which can be easily described as follows. For a given network, each node has supply/demand request and all the arcs between nodes have variable and fixed costs. The aim of the FCNF is to select the arcs and assign flow on them to transfer commodities from supply nodes to demand nodes,  such that the total cost is minimum. Many practical problems, such like transportation problem \citep{balinski1961fixed,el2013hybrid}, lot sizing problem \citep{steinberg1980optimal}, facility location problem \citep{nozick2001fixed}, network design \citep{lederer1998airline, ghamlouche2003cycle, costa2005survey} and others \citep{jarvis1978optimal, armacost2002composite, molla2011solving, zhang2016prediction, zhang2017bridge, nicholson2016optimal, zhang2018probabilistic, zhang2010lattice, zhang2016multi, zhang2010reformed, zhang2017resilience, zhang2016resilience} can be modeled as the FCNF. 

The FCNF is NP-hard \citep{GareyJ79} and over the decades, a significant number of papers have been published providing solution approaches to the FCNF.  In 1966, \citet{driebeek1966algorithm} proposed an algorithm to solve a mixed integer problem, which contains a large number of continuous variables and a few integer variables. He solved the problem without integer constraints first and search for the optimum integer solution. This work was among the pioneers  of mixed integer programming. Many techniques commonly utilities branch and bound (B\&B) to search the exact solution of the FCNF. \citet{kennington1976new} presented a new branch and bound algorithm to fixed charge transportation problem, which exploited the underlying transportation structure. \citet{barr1981new} revised B\&B to solve large scale and sparse fixed charge transportation problem. \citet{cabot1984some} discussed three properties that may improve B\&B algorithm. \citet{palekar1990branch} developed a stronger conditional penalty for the FCNF problem, which reduced the B\&B enumeration and solving time significantly. \citet{ortega2003branch} developed branch and cut system for uncapacitated FCNF problem, including a heuristic
for the dicut inequalities and branching and pruning rules. \citet{hewitt2010combining} obtained better solution using careful neighborhood search from arc-based formulation of the FCNF and improved the lower bound by linear relaxation of the path-based formulation. 

 Due to lacking the tight upper and lower bounds, the B\&B algorithm might be computational difficulty for the large-scale and complex problems. Hence, approximate solution approaches to find the near-optimal solution for the FCNF have generated considerable research interests. \citet{balinski1961fixed} formulated fixed cost transportation problem as integer programming and proposed a approximate method of solution. \citet{sun1998tabu} applied Tabu search algorithm to solve the FCNF problem, using recency based and frequency based memories, intermediate and long term memory processes. For local search, they used network simplex method. \citet{KimPardalos99} developed a dynamic slope scaling procedure, incorporating variable and fixed costs together as a new coefficient and solve linear programming problem iteratively. \citet{monteiro2011ant} proposed a hybrid ant colony optimization algorithm to combine two aspects of meta-heuristic search behavior, exploration and exploitation. Based on spanning tree and Pr{\"u}fer number representation, \citet{molla2011solving} proposed an artificial immune algorithm and genetic algorithm and discussed the proper values of parameters. 
 
 State-of-the-art MIP solvers combine a variety of cutting plane techniques, heuristics and the branch and bound algorithm to find the global optimal solution. All the modern MIP solvers use preprocessing/pre-solve methods to find a better upper bound by taking information from the original formulations, then the solvers are able to reduce the search space during B\&B procedure, which significantly accelerate the entire solving processes \citep{bix2000mip}. More details of the preprocessing techniques can be found in the books of \citet{nemhauser1988integer}, \citet{wolsey1998integer}, \citet{fugenschuh2005computational} and \citet{mahajan2010presolving}.

\section{Identification of Critical Components}
\label{ICC}

Modern societies are heavily dependent on many distributed systems, e.g. communication networks \citep{cohen2000resilience}, electric power transmission networks \citep{dobson2007complex}, transportation networks \citep{zheng2007clustering}, and all these belongs to network science area. A substantial body of work has been done to identify critical components of networks. 
In general,the removal or blockage of one or more critical links could have direct and serious economic consequences in terms of overall system performance \citep{bell2000game,smith2003characterization}. Therefore, we need to identify critical segments in the network, as specified by several national and transnational directives \citep{birchmeier2007systematic}.
The definition of criticality is always associated with a metric, such as travel time, maximum traffic flow, reliability and resilience, etc and the failure of critical components affect the system performance most. The term 'criticality' is used to qualify the role that the elements play with respect to the global properties of the whole network. 

Many literature employed topological approach to identify  critical components, which is capable of identifying the network edges and nodes whose failure can induce a severe structural damage to the network through the physical disconnection of its parts \citep{crucitti2005locating, bompard2009analysis}. 
However, focusing on network topology neglects other characteristics of the network.  \citet{eusgeld2009role} employed simulation technique to capture the dynamics of the operational scenarios involving the most vulnerable parts of the critical infrastructure combined with the analysis of network topology. \citet{bier2007methodology} proposed a computational efficient greedy algorithm, named max line interdiction algorithm, to identify the highest load line in the power system. 
\citet{dheenadayalu2004analysis} identified critical highway links using localized level-of-service measures such as the volume/capacity (V/C) ratio. \citet{scott2006network} developed a comprehensive Network Robustness Index for identifying critical components and network performance considering traffic flow, capacity and network connectivity. Through hypothetical networks, they proved that the Network Robustness Index yields different greater system-wide benefits with respect to travel time savings, than solution identified by V/C ratio.  

Besides network analysis and simulation approach, optimization is widely to identify critical components. \citet{zio2012identifying} proposed a multi-objective optimization model aimed at the maximization of the importance of the importance of the groups and minimization of their dimension. They applied the model to analyze the Italian high-voltage electrical transmission network. Similarly, \citet{shen2012exact} developed three objectives including the maximization of the number of connected components, minimization of the largest component size and maximization of the minimum cost required to reconnect the whole network, by deleting a group of nodes. 

In addition, \citet{demvsar2008identifying} proposed a dual graph model to identify critical locations in a spatial network in terms of vulnerability risk.
\citet{crucitti2004model} proved that the disruption of a single critical node is sufficient to decrease the efficiency or performance of the entire system.
\citet{barker2013resilience} involves two critical importance measures considering the adverse impact while blocking the link and positive impact if the link is not disrupted in the disaster. 

In summary, the critical index can be built in terms of different scenarios.  Removing the network component and quantifying the corresponding system-wide consequence is widely employed to build the component importance index regardless which technique used in the literature.

\section{Principal Goals and Thesis Outlines}
\label{Principal Goals and Thesis Outlines}

As the best of our knowledge, none of existing literature have developed models to analyze and predict which arcs are used in the optimal solution of the FCNF. This is the first paper that employs statistical learning technique to analyze the optimal flow of the FCNF and distinguishes the features to predict the optimal arcs.  This thesis involves three tasks. \textbf{Task 1} will extract network features that can be used as predictors and how to develop a predictive model with these features. By solving thousands of random generated FCNF instances, we collect over $60,000$ observations and develop logistic regression model based on the dataset, which can be used to quantify the influence of each statistical significant network characteristics. \textbf{Task 2} will validate the predictive model through a series of diagnostic techniques will interpret the model. \textbf{Task 3} of this thesis, will discuss two applications of this model. The first one is to identify critical arcs in a network with respect to the FCNF problem without solving any optimization problems. The other one is named regression-based relaxation (RBR), a approximate method for the FCNF. 

The remainder of this thesis is organized as follows. Chapter \ref{Background} introduces the background of the FCNF and the logistic regression model. The whole process for developing logistic regression model is discussed in Chapter \ref{Optimal Flow Analysis} and Chapter \ref{Validation and Interpretation} validates and interprets the predictive model. Chapter \ref{Applications} introduces two applications of the predictive model. Chapter \ref{conclusion} and \ref{future} summarize the thesis and future work, respectively.

\chapter{Background}
\label{Background}

\section{Fixed Charge Network Flow Formulation}
\label{Fixed Charge Network Flow Formulation}
The fixed charge network flow (FCNF) problem is described on a network $G = (N,A)$, where $N$ and  $A$ are the set of node index $i$ and arc index $(i,j)$, respectively.  Let $c_{ij}$ and $f_{ij}$ denote the variable and fixed cost of arc $(i,j) \in A$, respectively. Each node has a supply/demand request $R_{i}$ ($R_{i}>0$ if node $i$ is a supply node; $R_{i}<0$ if node $i$ is a demand node; otherwise, $R_{i}=0$). Artificial capacity, $M_{ij}$, is used in the problem formulation to ensure that the fixed cost $f_{ij}$ is incurred whenever there is a positive flow on arc $(i,j) \in A$. There are two variables in the FCNF,  $x_{ij}$ denotes the flow and $y_{ij}$ represents the corresponding decision variable on arc $(i,j) \in A$ . All the instances in this work are  single-commodity FCNF and the formulation is as follows,

\begin{align} 
\text{min} & \sum_{(i,j) \in A} (c_{ij}x_{ij} + f_{ij}y_{ij})  \label{obj} \\
\text{s.t.} & \sum_{(i,j) \in A} x_{ij} - \sum_{(j,i) \in A} x_{ji} = R_{i}&
\forall{i \in N} \label{equilibrium} \\
&0 \leq x_{ij} \leq M_{ij}y_{ij}&
\forall{(i,j) \in A} \label{bigM} \\
&y_{ij} \in \lbrace 0,1 \rbrace&
\forall{(i,j) \in A} \label{binary}
\end{align}
Here, the objective function in (\ref{obj}) is nonlinear. Constraint (\ref{equilibrium}) ensures that the inflow and outflow satisfy the supply/demand at node $i \in N$. 
$M_{i,j}$ is arc capacity in constraint (\ref{bigM}) to ensure that the flow on arc $(i,j)\in A$ can be positive only when the arc $(i,j)\in A$ is open ($y_{ij}=1$). If arc$(i,j)$ is not capacitated, $M_{ij}$ should be artificially set to a number which is larger than the total supply requests to not inhibit feasible solution.  Constraint (\ref{binary}) defines $y_{ij}$ as binary, which makes the problem a 0-1 mixed integer programming problem.

\section{Logistic Regression Model}
\label{Logistic Regression Model}

The logistic regression is a classification model and only has one response/dependent variable. The logistic regression is commonly used in data mining applications, computer science, biology and so on \citep{camdeviren2007comparison}. There are two classes of the logistic regression, binomial and multinomial. The response variable of binomial logistic regression has exactly two classes or binary outcomes, while multinomial logistic regression has more than two. The dependent variable in this paper, denoted by $Y$, is binary and
\begin{align*}
Y = \begin{cases}
1, \text{arc is used in the FCNF}\\
0, \text{otherwise}.
\end{cases}
\end{align*} 
The initial output of the logistic regression is not $0$ or $1$, but the probability that the response variable equals to $1$, which in this work refers to the likelihood that the arc has positive flow in the optimal solution of the FCNF. The probability is derived by a logistic function of a variety of independent predictors and the logistic regression coefficients. We use $\pi(p) = P(Y=1|p_1,p_2,\dots,p_k,\dots, p_K)$  to represent the conditional probability of $Y=1$ given $p_{k}$ when the logistic regression is used. The specific logistic regression model is, 
\begin{align}
\pi(p) = \dfrac{1}{1+e^{-(\beta_0+\beta_1p_1+\beta_2p_2+\dots+\beta_kp_k+\dots+\beta_Kp_K)}} \label{logisticregression}
\end{align} 
where $K$ is the total number of independent predictors, $p_{k}$ is the $k^{\text{k}}$ predictor and $\beta_{k}$ is the $k^{\text{th}}$  parameter. 
Let $g(p)$ denote the logistic transformation of $\pi(p)$ and defined as,
\begin{align*}
g(p) & = ln[\dfrac{\pi(p)}{1-\pi(p)}]\\
& = \beta_0+\beta_1p_1+\beta_2p_2+\dots+\beta_kp_k+\dots+\beta_Kp_K
\end{align*}
The logit, $g(p)$, is linear in its parameters. The logistic transformation has many desirable properties of linear regression which we will use to discuss the model and interpretation. 
Transforming the likelihood to binary value needs a cut-off point, above which the response variable is set to $1$. The details of setting the cut-off value is discussed in Chapter \ref{Validation and Interpretation}.

\chapter{Optimal Flow Analysis}
\label{Optimal Flow Analysis}

\section{Network Features Extraction}
\label{Network Features Extraction}
In this section, we extract features from network to predict which arcs will be open in the optimal solutions of the FCNF with four types of features (network, arcs, relaxation solutions and nodes). In the network level, the features include the total number of nodes ($n$), the number of arcs ($m$), the sum of supply requests ($S$) and the density of network  , which is denoted by $\rho = \frac{m}{2\binom{n}{2}}$. The total supplies is transformed to the average supply per node, $\bar{S}=\frac{S}{n}$. 
Arc $(i,j)\in A$ has variable cost $c_{ij}$ and fixed cost $f_{ij}$ based on formulation (\ref{obj}). Let $\gamma_{ij}$ denote the ratio between fixed cost and variable cost of arc $(i,j) \in A$,
\begin{align*}
\gamma_{ij} = \frac{f_{ij}}{c_{ij}} \qquad \forall{(i,j) \in A}.
\end{align*}
Let $l_{ij}$ denote the flow solution of arc $(i,j) \in A$ in linear relaxed FCNF problem and the normalized value of  $l_{ij}$ is denoted by $\bar{l}_{ij}$,
\begin{align*}
\bar{l}_{ij}=\frac{l_{ij}}{S} \qquad \forall (i,j) \in A.
\end{align*}
The value of $\bar{l}_{ij}$ indicates the optimal flow of the linearized FCNF, and another binary predictor of the optimal arc is created, which is denoted by $l^B_{ij}$, 
\begin{align*}
l^B_{ij} = \begin{cases}
1, \bar{l}_{ij} > 0\\
0, \text{otherwise}.
\end{cases}
\end{align*}
The two predictors are both related with the optimal flow in the linear relaxed FCNF, but reflect two different types of information. The value of $\bar{l}_{ij}$ indicates how much flow on arc $(i,j) \in A$ and $l^B_{ij}$ shows the arc is open or not in the linearized FCNF.

The information of head node $i$ and tail node $j$ of arc $(i,j) \in A$ should also be considered as the features that may facilitate the predictive model due to the fact that the goal of FCNF is to transfer commodities from supply nodes to demand nodes through several transshipment nodes. Accordingly, we detail the predictors associated with head node $i \in N$, with which tail node $j \in N$ is same. Let $t_{i}$ denote the type of node $i \in N$, 
\begin{align*}
t_{i} = 
\begin{cases} 1, r_{i}>0 \\
0, r_{i} = 0 \\
-1, r_{i} < 0\end{cases}.
\end{align*} 
This is to say if node $i$ is a supply node, $t_{i}=1$; if node $i$ is a transshipment node, $t_{i}=0$ ; and otherwise if node $i$ is a demand node, $t_{i}=-1$. Let $d_{i*}$ denote the outdegree of node $i$. It is reasonable to assume that the nodes adjacent to node $i$ have a significant influence on the corresponding response variable. For example, if a supply node is connected with node $i$, the probabilities of this arc and other arcs out from node $i$ to be selected will increase. Let $d_{i*}^S$ ($d_{i*}^D$) denote the number of supply (demand) tail nodes adjacent to node $i$. If a demand node is connected with node $i$, the probabilities of this arc and other arcs into node $i$ to be selected will increase as well. Let $r_{i*}^S$ ($r_{i*}^D$) denote the sum of supply (demand) requirements of the tail endpoints adjacent to node $i$. In the same way, the indegree of node $i$ is denoted as $d_{*i}$, the number of supply (demand) head nodes adjacent to node $i$ is denoted by $d_{*i}^S$ ($d_{*i}^D$), and sum of supply (demand) requests of head endpoints adjacent to node $i$ are denoted as $r_{*i}^S$ ($r_{*i}^D$). Figure \ref{nodeexample} illustrates the notations with a small directed network, in which the number under each node is the requirements of that node and arrow shows the direction of the corresponding arc. The value of $d_{i}$ is 6 because we have three arcs into node $i$ and three arcs out from node $i$. The requirement of node $i$ is $100$ and thus the value of $t_{i}$ is $1$. All the notations defined before are showed in this figure.
\myfigure{
	\centering
	\includegraphics[width=\textwidth]{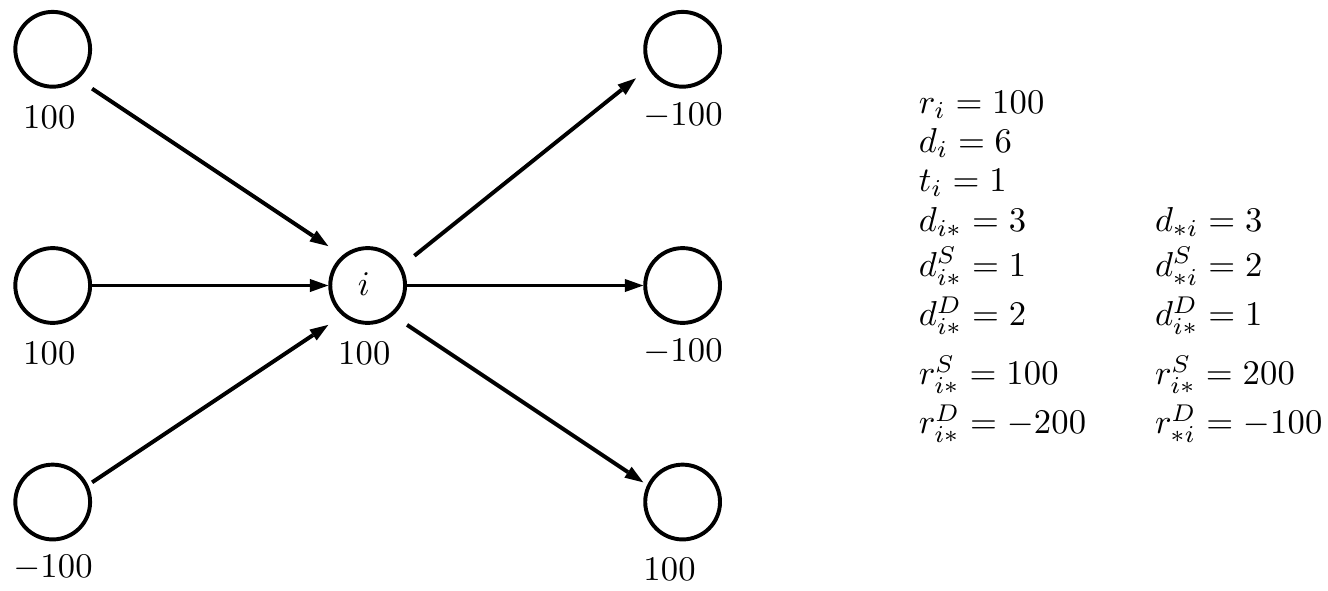}
}{Example of Node $i$}{nodeexample}

We observe that all the predictors with respect to degrees and requirements are correlated with the size and total requirements of each specified instance, consequently it is necessary to normalize such predictors. All the predictors with respect to degrees are divided by the number of nodes $N$, and then we have $\bar{d}_{i}$, $\bar{d}_{i*}$, $\bar{d}_{i*}^S$ , $\bar{d}_{i*}^D$, $\bar{d}_{*i}$, $\bar{d}_{*i}^S$, $\bar{d}_{*i}^D$ ( $\forall i \in N$). All the predictors related with requirements are divided by total supply $S$, and we get $\bar{r}_{i}$, $\bar{r}_{i*}^S$, $\bar{r}_{i*}^D$, $\bar{r}_{*i}^S$, $\bar{r}_{*i}^D$ ($\forall i \in N$). In summary, we totally have $33$ predictors for a individual arc $(i,j) \in A$ as summarized in Table \ref{featurelist}.

\mytable{
	\begin{center}
		\begin{tabular}[!ht]{c  l}
			\hline
			Notation & Description \\
			\hline
			$n$	&	number of nodes	\\
			$m$ & total number of arcs \\
			$\rho$	&	network density	\\
			$\bar{S}$	&	average supply	\\
			$c_{ij}$	&	variable cost of arc  $(i,j)$	\\
			$f_{ij}$	&	fix cost of arc $(i,j)$	\\
			$\gamma_{ij}$	&	ratio between fix cost and variable cost of arc $(i,j)$	\\
			$\bar{l}_{ij}$	&	normalized value of linearized relaxation solution of arc $(i,j)$	\\
			$l^B_{ij}$ & binary version of linearized relaxation solution of arc $(i,j)$	\\
			$t_{i}$	&	type of head node $i$	\\
			$t_{j}$	&	type of tail node $j$	\\
			$\bar{r}_{i}$	&	requirements of head node $i$	\\
			$\bar{r}_{j}$	&	requirements of tail node $j$	\\
			$\bar{r}_{i*}^S$ & sum of supply requirements of tail endpoints adjacent to node $i$ \\
			$\bar{r}_{i*}^D$	& sum of demand requirements of tail endpoints adjacent to node $i$	\\
			$\bar{r}_{*i}^S$	&	sum of supply requirements of head endpoints adjacent to node $i$	\\
			$\bar{r}_{*i}^D$	&	sum of demand requirements of head endpoints adjacent to node $i$	\\
			$\bar{r}_{j*}^S$	&	sum of supply requirements of tail endpoints adjacent to node $j$	\\
			$\bar{r}_{j*}^D$	&	sum of demand requirements of tail endpoints adjacent to node $j$	\\
			$\bar{r}_{*j}^S$	&	sum of supply requirements of head endpoints adjacent to node $j$	\\
			$\bar{r}_{*j}^D$	&	sum of demand requirements of head endpoints adjacent to node $j$	\\
			$\bar{d}_{i*}$	&	outdegree of head node $i$	\\
			$\bar{d}_{i*}^S$	& number of supply tail nodes adjacent to node $i$	\\
			$\bar{d}_{i*}^D$	&	number of demand tail nodes adjacent to node $i$	\\
			$\bar{d}_{*i}$	&	indegree of head node $i$	\\
			$\bar{d}_{*i}^S$	& number of supply head nodes adjacent to node $i$	\\
			$\bar{d}_{*i}^D$	&	number of demand head nodes adjacent to node $i$	\\
			$\bar{d}_{j*}$	&	outdegree of tail node $j$	\\
			$\bar{d}_{j*}^S$	&	number of supply tail nodes adjacent to node $j$	\\
			$\bar{d}_{j*}^D$	&		number of demand tail nodes adjacent to node $j$	\\
			$\bar{d}_{*j}$	&	indegree of tail node $j$	\\
			$\bar{d}_{*j}^S$	&	number of supply head nodes adjacent to node $j$	\\
			$\bar{d}_{*j}^D$	&	number of demand head nodes adjacent to node $j$	\\
			\hline
		\end{tabular}
	\end{center}}{Entire Feature List of arc $(i,j) \in A$ } {featurelist}

\section{Data Collection}
\label{Data Collection}

In order to obtain sufficient data set, we create $1249$ single-commodity FCNF instances randomly, each which has characteristics corresponding to difficult FCNF problem instances (e.g. high fixed to variable cost ratio). All problems are solved by GUROBI 5.6 on the platform of Windows 7 64bit machine with Intel Xeon CPU E5-1620 and 8 GB RAM. The tests include network works with range in $5$ to $15$, thus we can solve them optimally quickly. For each problem, the number of arcs $m$ is randomly selected. Specifically, we randomly choose $n-1 \leq \dfrac{m}{2} \leq \dfrac{n(n-1)}{2}$ and create a connected network instance where each of the $\dfrac{m}{2}$ undirected arcs is replaced by two directed arcs. The percentage of supply, demand, and transshipment nodes are respectively randomly selected with approximate probabilities $0.2$, $0.2$ and $0.6$. The probabilities are approximate in that adjustments are made to ensure an instance is feasible. The variable costs and fixed costs for each link are randomly assigned on $U (0, 20)$ and $U (20000, 60000)$, respectively. The total requirements for each supply node is randomly assigned on $U(1000,2000)$. The total requirements of supply node is distributed randomly as negative requirements to the demand nodes. 
The statistics information of the final predictors created in Table \ref{featurelist} in our train dataset are reported in Table \ref{statis}. Please note that head nodes and tail nodes share the same statistical information.

\mytable{
	\begin{tabular}{c | c c c c c c}
		\hline
		Notation	&	Min.   	&	1st Qu.	&	Median 	&	Mean   	&	3rd Qu.	&	Max.   	\\
		\hline
		$n$	&	5.00	&	10.00	&	12.00	&	11.71	&	14.00	&	15.00	\\
		$m$ & 8.00 &  50.00  & 84.00 &  89.86 & 126.00 & 210.00 \\
		$\rho$	&	0.13	&	0.51	&	0.72	&	0.69	&	0.88	&	1.00	\\
		$\bar{S}$	&	84.93	&	307.20	&	456.75	&	480.50	&	622.93	&	1325.80	\\
		$c_{ij}$	&	0.00	&	2.52	&	5.00	&	5.01	&	7.49	&	10.00	\\
		$f_{ij}$	&	20000	&	29998	&	39983	&	40024	&	50056	&	60000	\\
		$\gamma_{ij}$	&	2018	&	5242	&	7997	&	42839	&	15942	&	141547369	\\
		$\bar{l}_{ij}$	&	0.00	&	0.00	&	0.00	&	0.02	&	0.00	&	1.00	\\
		${l}^B_{ij}$	&	0.00	&	0.00	&	0.00	&	0.11	&	0.00	&	1.00	\\
		$t_{i}$, $t_{j}$	&	-1.00	&	-1.00	&	0.00	&	0.15	&	1.00	&	1.00	\\
		$\bar{r}_{i}$, $\bar{r}_{j}$	&	-1.00	&	-0.16	&	0.00	&	0.00	&	0.15	&	1.00	\\
		$\bar{r}_{i*}^S$, $\bar{r}_{j*}^S$	&	0.00	&	0.40	&	0.60	&	0.60	&	0.90	&	1.00	\\
		$\bar{r}_{i*}^D$, $\bar{r}_{j*}^S$	&	-1.00	&	-1.00	&	-0.69	&	-0.65	&	-0.45	&	0.00	\\
		$\bar{r}_{*i}^S$, $\bar{r}_{*j}^S$	&	0.00	&	0.00	&	0.00	&	0.64	&	1.01	&	10.00	\\
		$\bar{r}_{*i}^D$, $\bar{r}_{*j}^D$	&	-11.00	&	-1.00	&	0.00	&	-0.66	&	0.00	&	0.00	\\
		$\bar{d}_{i*}$, $\bar{d}_{j*}$	&	0.07	&	0.50	&	0.67	&	0.64	&	0.83	&	0.93	\\
		$\bar{d}_{i*}^S$, $\bar{d}_{j*}^S$	&	0.00	&	0.17	&	0.27	&	0.28	&	0.38	&	0.83	\\
		$\bar{d}_{i*}^D$, $\bar{d}_{j*}^D$	&	0.00	&	0.10	&	0.17	&	0.18	&	0.25	&	0.67	\\
		$\bar{d}_{*i}$, $\bar{d}_{*j}$	&	0.07	&	0.50	&	0.67	&	0.64	&	0.83	&	0.93	\\
		$\bar{d}_{*i}^S$, $\bar{d}_{*j}^S$	&	0.00	&	0.00	&	0.00	&	0.28	&	0.62	&	0.93	\\
		$\bar{d}_{*,i}^D$, $\bar{d}_{*,j}^D$	&	0.00	&	0.00	&	0.00	&	0.18	&	0.36	&	0.93	\\
		\hline
		$y_{ij}$	&	0.00	&	0.00	&	0.00	&	0.11	&	0.00	&	1.00	\\
		\hline
	\end{tabular}
}{Statistic Information of arc $(i,j) \in A$ of Train Dataset} {statis}

Every arc in each instance is an unique record in our dataset and there are $61594$ rows in train dataset with $1067$ instances and $37651$ rows in test dataset with $182$ instances. The $y_{ij}$ is the binary variable in the optimal solution of the FCNF instances and it is apparently to observe from Table \ref{statis} that the average value of $y_{ij}$ is quite low, which means only $11\%$ of arcs are used in the optimal solutions overall and this value is smaller when the problem is larger. \citet{zhang2014OSEA} provides evidence to illustrate that only a tiny proportion of integer variables are nonzero in the optimal solution of general mixed integer programming problems. Among the $61594$ records, only $6755$ arcs are used ($y_{ij}=1$) and a huge proportion of arcs are relative useless ($y_{ij}=0$). Accordingly, the dataset now is biased and can not be used to perform analysis at this point. Moreover, the records are related with each other in the same instance since an trivial arc could be chosen if the optimal arc is removed, which is like an underlying feature affecting the probability, consequently it is necessary to adjust the data set to weaken this effect.
To make the dataset fair and reasonable to train predictive model, we use the undersampling technique to adjust train dataset as described below. Reversely, we can also use oversampling to increase the number of records where  $y_{ij}=1$. 

\noindent \textit{Process of undersampling:}\\
\textit{Step 1. Fetch the records that $y_{ij}$ equals $1$ and number of rows is stored as $Row1$;\\
	Step 2. Set $Row0=0$ as the number of rows that $y_{ij}$ equals $0$;\\
	Step 3. Randomly fetch a record from the records that $y_{ij}$ equals $0$, and $Row0 = Row0 + 1$;\\
	Step 4. If $Row0 < Row1$, go to step 3; otherwise, output all records and stop.}

Following above process, the final train dataset contains $13349$ rows and the mean value of $y_{ij}$ is $0.51$. Note, the undersampling procedure is only performed on train dataset but not on the test dataset.

\section{Akaike Information Criterion Selection}
\label{AICselection}

In this subsection, we use stepwise variable selection algorithm to select the features based on the Akaike information criterion ($AIC$) \citep{akaike1974new}. The formulation of $AIC$ is \citep{venables2002modern},
\begin{align}
AIC = -2log L + 2K, \label{AIC}
\end{align}
where $L$ is the likelihood and $K$ is the number of predictors. From the formulation (\ref{AIC}), it is straightforward to identify that step wise regression process handles the trade-off between the performance and the complexity of the model.  

Step wise $AIC$ process could be performed with three directions: backward, forward and bi-direction. In this work, the process of backward stepwise selection starts from $33$ predictors and the value of $AIC$ equals to $7,924.72$, drops one feature at each step, and finally stops at $26$ predictors with $AIC$ value of $7,910$. The final logistic regression model is reported in Table \ref{logistic}, in which the value of coefficients ($\beta$) indicates the magnitude and direction of the features affecting the probability that an arc is in the optimal solution. The odds ratio (OR) is the exponential of the coefficients ($\beta$) and can be derived by,
\begin{align*}
\pi(p_{k}) & = \frac{1}{1+e^{-\beta_{0}+\beta_{k}p_{k}}}, \\
odds(\pi(p_{k})) & = \frac{\pi(p_{k})}{1-\pi(p_{k})}, \\
OR(p_{k}) & = \frac{odds(\pi(p_{k}+1))}{odds(\pi(p_{k}))} = e^{\beta_{k}}.
\end{align*}
The value of OR provides an interpretation for the coefficients $\beta$, which is the change of dependent variable for 1-unit increase/decrease of the corresponding independent predictor. If the predictor is binary, replace the $\pi(p_{k}+1)$ with $\pi(p_{k}=1)$ and  $\pi(p_{k})$ with $\pi(p_{k}=0)$. 

\mytable{
	\begin{tabular}{c | c c c c c}
		\hline
		Predictors	&	$\beta$	&	Std. Error	&	OR	&	Pr($>|z|$)	&	Significant code	\\
		\hline
		
		(Intercept)		&	8.32E+00	&	4.41E-01	&	4121.613533	&	$<$2.00E-16	&	***	\\
		$n$		&	-4.80E-02	&	2.63E-02	&	0.953095662	&	0.067393	&	.	\\
		$m$		&	-5.52E-03	&	1.86E-03	&	0.994499185	&	0.003003	&	**	\\
		$c_{ij}$		&	-8.79E-02	&	1.04E-02	&	0.915861616	&	$<$2.00E-16	&	***	\\
		$f_{ij}$		&	-1.50E-04	&	3.49E-06	&	0.999850511	&	$<$2.00E-16	&	***	\\
		$\gamma_{ij}$	&	1.28E-07	&	1.15E-07	&	1.000000128	&	0.268276	&		\\
		$t_{i}=0$	&	-5.43E-01	&	3.15E-01	&	0.58094453	&	0.084507	&	.	\\
		$t_{i}=1$		&	2.17E+00	&	2.60E-01	&	8.793387337	&	$<$2.00E-16	&	***	\\
		$t_{j}=0$		&	-3.01E+00	&	2.98E-01	&	0.049242412	&	$<$2.00E-16	&	***	\\
		$t_{j}=1$		&	-2.38E+00	&	2.10E-01	&	0.092458073	&	$<$2.00E-16	&	***	\\
		$\bar{l}^B_{ij}$		&	1.40E+00	&	1.16E-01	&	4.067383833	&	$<$2.00E-16	&	***	\\
		$\bar{l}_{ij}$		&	5.76E+00	&	5.11E-01	&	318.3018034	&	$<$2.00E-16	&	***	\\
		$\bar{r}_{i}$	&	9.67E-01	&	3.38E-01	&	2.629516529	&	0.004193	&	**	\\
		$\bar{r}_{i*}^S$	&	7.65E-01	&	1.73E-01	&	2.148779486	&	9.98E-06	&	***	\\
		$\bar{r}_{i*}^D$	&	8.78E-01	&	1.99E-01	&	2.406323346	&	9.82E-06	&	***	\\
		$\bar{r}_{*i}^S$	&	-2.15E-01	&	5.46E-02	&	0.806380148	&	8.15E-05	&	***	\\
		$\bar{r}_{*i}^D$	&	4.64E-01	&	8.81E-02	&	1.590104918	&	1.40E-07	&	***	\\
		$\bar{r}_{j*}^S$	&	-7.97E-01	&	1.67E-01	&	0.450724045	&	1.86E-06	&	***	\\
		$\bar{r}_{j*}^D$	&	-9.17E-01	&	2.04E-01	&	0.399596496	&	6.71E-06	&	***	\\
		$\bar{r}_{*j}^S$		&	-1.22E-01	&	4.16E-02	&	0.884971357	&	0.003284	&	**	\\
		$\bar{d}_{i*}$		&	-2.59E+00	&	5.67E-01	&	0.07487015	&	4.79E-06	&	***	\\
		$\bar{d}_{*i}$		&	-2.69E+00	&	3.37E-01	&	0.067609958	&	1.37E-15	&	***	\\
		$\bar{d}_{i*}^D$	&	-1.30E+00	&	7.08E-01	&	0.272259397	&	0.066353	&	.	\\
		$\bar{d}_{*i}^S$		&	1.84E+00	&	4.98E-01	&	6.296538261	&	0.000221	&	***	\\
		$\bar{d}_{*i}^D$	&	4.57E+00	&	5.85E-01	&	96.73739121	&	5.60E-15	&	***	\\
		$\bar{d}_{j*}$		&	-2.30E+00	&	5.69E-01	&	0.100258844	&	5.19E-05	&	***	\\
		$\bar{d}_{j*}^D$		&	-1.58E+00	&	7.49E-01	&	0.205769226	&	0.034836	&	*	\\
		$\bar{d}_{*j}^S$	&	3.19E+00	&	4.96E-01	&	24.21567135	&	1.26E-10	&	***	\\
		$\bar{d}_{*j}^D$	&	1.26E+00	&	5.04E-01	&	3.536013632	&	0.01215	&	*	\\
		\hline
	\end{tabular}
	\begin{tablenotes}
		\small
		\item Significant codes:  `***', 0.001; `**', 0.01; `*', 0.05; 
		\item $i$ referred to the head node $i$ and $j$ referred to the tail node $j$ of arc $(i,j) \in A$
	\end{tablenotes}
}{Logistic Regression Model with $26$ predictors} {logistic}

\chapter{Validation and Interpretation}
\label{Validation and Interpretation}

\section{Model Diagnostics}
\label{Diagnostics}

In this subsection, a series of diagnostic techniques are performed to validate the logistic regression model as showed in Table \ref{logistic}. Firstly, k-fold cross validation is used on the train data set. Specifically, k equals to $10$ in our test, which is the most common used value of k \citep{mclachlan2005analyzing}. The train data set is randomly partitioned into $10$ equal size subsamples. A single subsample is retained as validation data set for testing the model and the remaining $9$ subsamples are used as training data. This process is repeated $10$ times, in which each subsample is employed as test exactly once. The $10$ results then are averaged to give one estimation. In this test, the average estimation accuracy is $0.884$. 
 
Then, analysis of confusion matrix,  also named contingency table or an error matrix \citep{stehman1997selecting}, is performed on the train dataset as well, which is widely used to visualize the performance of logistic regression. In the confusion matrix, each row represents the number of $1$ or $0$ from observations ($y_{ij}$ in Table \ref{statis}), which is called actual value, and every column shows the predict values from logistic regression model. Confusion matrix reports four derivations, false positives (FP), false negatives (FN), true positives (TP), and true negatives (TN). Let $FPR$ denote the false positive rate and $FNR$ denote the false negative rate. The formulation to compute $FPR$ and $FNR$ are,
\begin{align*}
FPR = \frac{FP}{FP + TN},  \\ 
FNR = \frac{FN}{FN + TP}.  
\end{align*}
As explained before, the logistic regression model only gives the probability that an arc is used in the  optimal solution. Therefore, there is a need to find a cut-off point, so that the response variable with the probability larger than the point is set to $1$; otherwise, the response variable is assigned with $0$. To find the optimal cut-off point, two costs are associated with  $FPR$ and $FNR$, represented by $C_{FPR}$ and $C_{FNR}$, and the false cost function is, 
\begin{align}
C_{FPR}FPR + C_{FNR}FNR.  \label{cost}
\end{align}
The optimal cut-off point is obtained by solving the minimization problem with  the false total cost based on formulation (\ref{cost}) as objective function,. Accordingly, the optimal solution varies with different cost structures, which should be determined by specified applications. Assume the cost of $FPR$ equals to $FNR$ in this thesis,
Figure \ref{costcutoff} plots the cost as a function of cut-off point and the optimal cut-off point is $0.49$.  The detailed confusion matrix with cut-off point equals to $0.49$ is listed in Table \ref{CM}. 
\myfigure{
	\centering
	\includegraphics[width=0.8\textwidth]{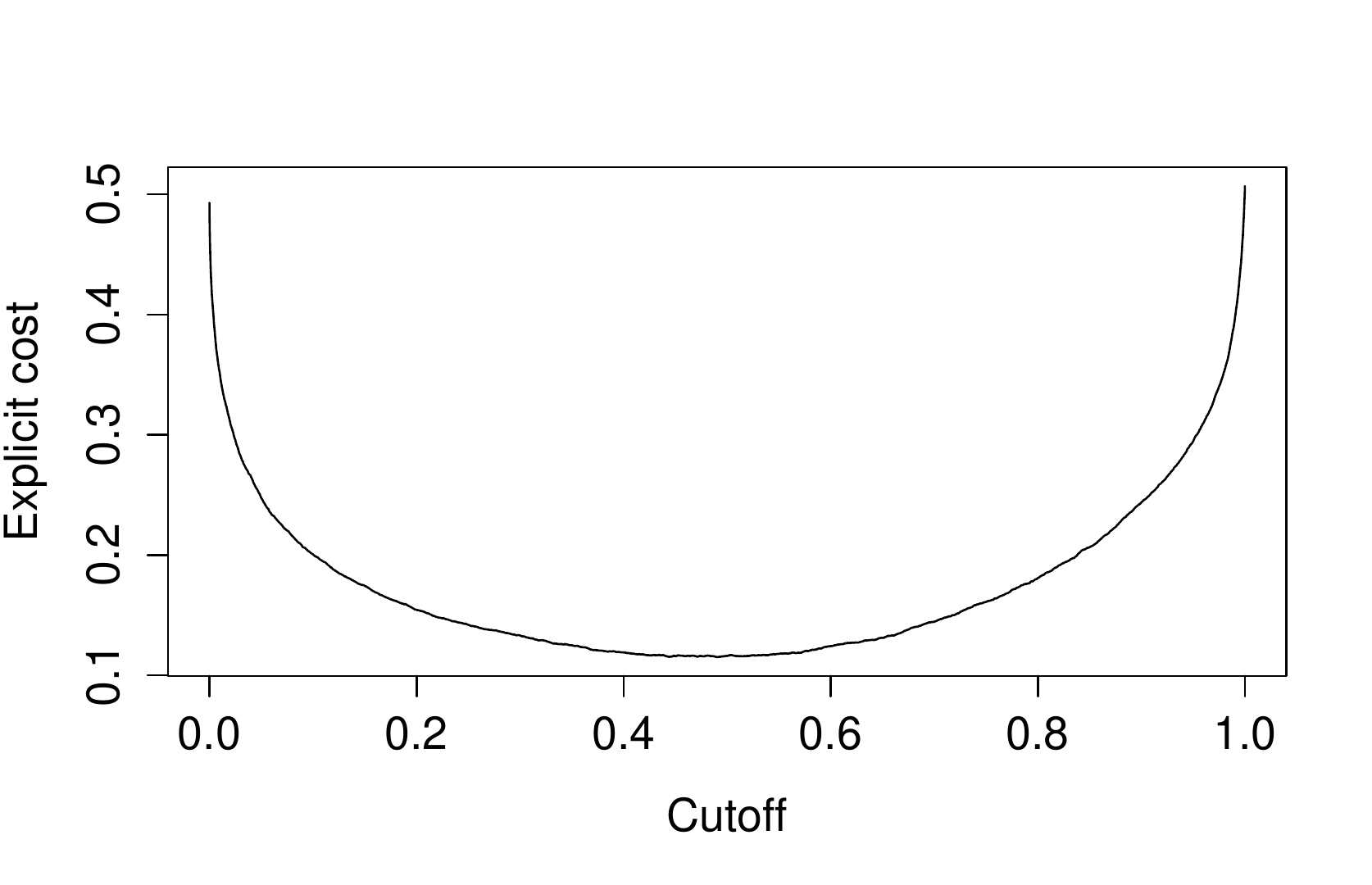}
}{Cost Value over Cut-off of Train Dataset}{costcutoff}

\mytable{
	\begin{tabular}{c   c | c c }
		\cline{3-4}
		&  & \multicolumn{2}{ c| }{Predictive} \\ \cline{3-4}
		\multicolumn{1}{c }{} & & False & 
		\multicolumn{1}{ c|}{True}\\  \cline{1-4}
		\multicolumn{1}{ |c|}{\multirow{2}{*}{Actual}} & 0 & 5683 (TN) & 
		\multicolumn{1}{ c|}{899 (FN)} \\
		\multicolumn{1}{ |c|}{} & 1 &  638 (FP) & 
		\multicolumn{1}{ c|}{6129 (TP)} \\
		\hline
	\end{tabular}
}{Confusion Matrix with Cut-off $0.49$ of Train Dataset} {CM}

\myfigure{
	\centering
	\includegraphics[width=0.8\textwidth]{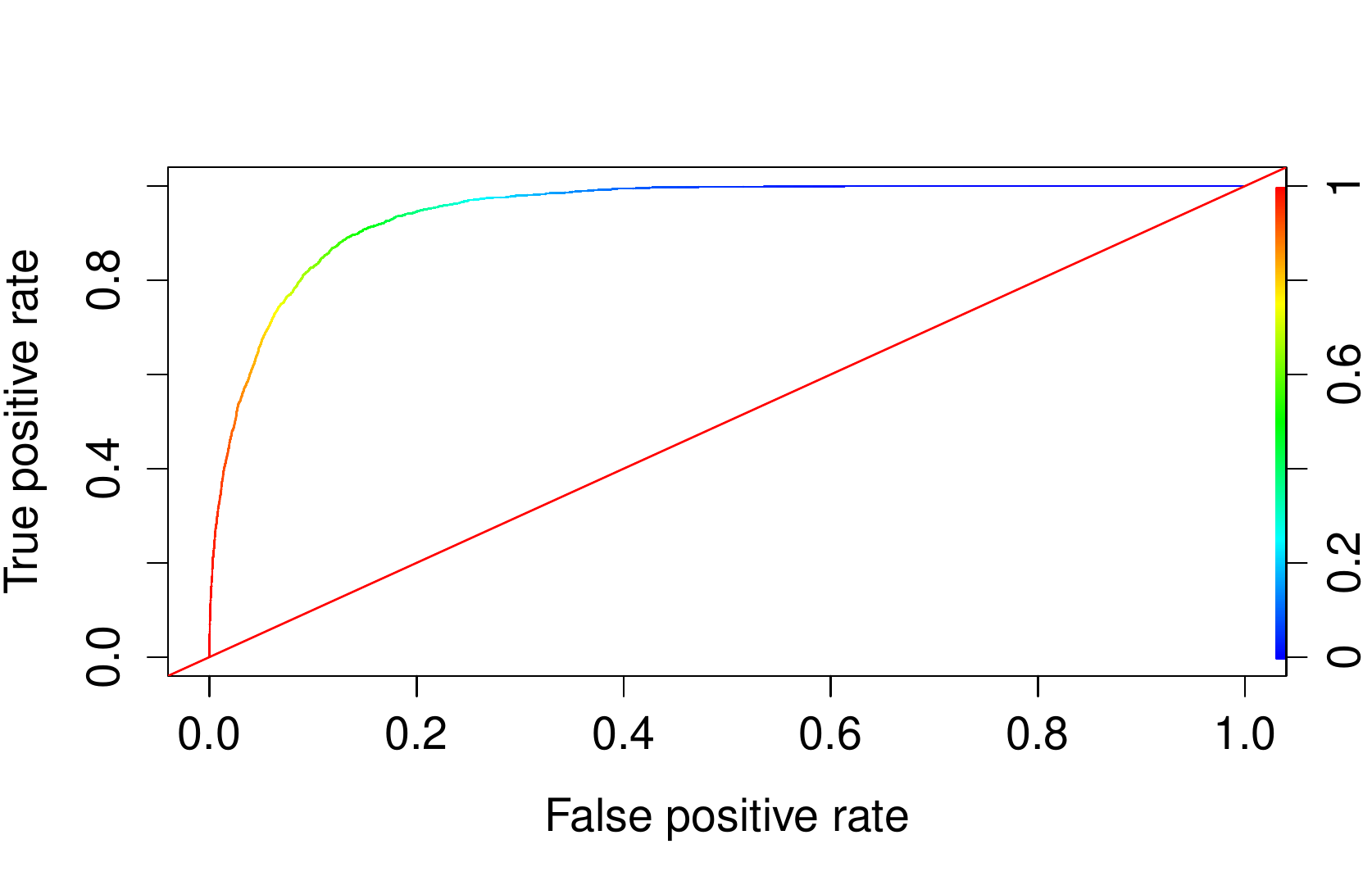}
}{ROC Curve on Test Dataset}{roc}

\mytable{
	\begin{tabular}{c | c c c c c c}
		
		\hline
		Notation	&	Min.   	&	1st Qu.	&	Median 	&	Mean   	&	3rd Qu.	&	Max.   	\\
		\hline
		$n$	&	15.00	&	19.00	&	20.00	&	20.62	&	23.00	&	25.00	\\
		$m$ & 8.00 &  50.00 &  84.00 &  89.86 & 126.00 & 210.00 \\
		$\frac{1}{N^2}$	&	0.00	&	0.00	&	0.00	&	0.00	&	0.00	&	0.00	\\
		$\bar{S}$	&	58.22	&	312.20	&	476.90	&	464.20	&	565.50	&	993.60	\\
		$c_{ij}$	&	0.00	&	2.50	&	5.00	&	5.01	&	7.52	&	10.00	\\
		$f_{ij}$	&	20000.00	&	30150.00	&	40260.00	&	40140.00	&	50180.00	&	60000.00	\\
		$\gamma_{ij}$	&	2024.00	&	5256.00	&	8038.00	&	42790.00	&	15940.00	&	78880000.00	\\
		$\bar{l}_{ij}$	&	0.00	&	0.00	&	0.00	&	0.01	&	0.00	&	0.99	\\
		$t_{i}$	&	-1.00	&	-1.00	&	0.00	&	0.15	&	1.00	&	1.00	\\
		$\bar{r}_{i}$	&	-1.00	&	-0.10	&	0.00	&	0.00	&	0.09	&	0.99	\\
		$\bar{r}_{i*}^S$	&	0.00	&	0.47	&	0.68	&	0.65	&	0.86	&	1.00	\\
		$\bar{r}_{i*}^D$	&	-1.00	&	-0.88	&	-0.68	&	-0.65	&	-0.46	&	0.00	\\
		$\bar{r}_{*i}^S$	&	0.00	&	0.00	&	0.00	&	0.65	&	1.09	&	16.11	\\
		$\bar{r}_{*i}^D$	&	-19.00	&	-0.98	&	0.00	&	-0.66	&	0.00	&	0.00	\\
		$\bar{d}_{i*}$	&	0.04	&	0.50	&	0.68	&	0.65	&	0.83	&	0.96	\\
		$\bar{d}_{i*}^S$	&	0.00	&	0.19	&	0.27	&	0.27	&	0.36	&	0.61	\\
		$\bar{d}_{i*}^D$	&	0.00	&	0.10	&	0.18	&	0.18	&	0.25	&	0.48	\\
		$\bar{d}_{*i}$	&	0.04	&	0.50	&	0.68	&	0.65	&	0.83	&	0.96	\\
		$\bar{d}_{*i}^S$	&	0.00	&	0.00	&	0.00	&	0.27	&	0.60	&	0.96	\\
		$\bar{d}_{*,i}^D$	&	0.00	&	0.00	&	0.00	&	0.19	&	0.35	&	0.96	\\
		\hline												
		$y_{ij}$	&	0.00	&	0.00	&	0.00	&	0.07	&	0.00	&	1.00	\\
		\hline
	\end{tabular}
}{Static Information of Selected Features of Test Dataset} {teststatis}

Furthermore, the model is validated on test data set, which is never touched before. The statistics information of test data set is summarized in Table \ref{teststatis}, in which we can see the number of nodes ranges between $15$ and $25$. Therefore, the problems solved in the test data set are larger than train data and none of problems exist in both data sets. Firstly, Table \ref{CMT} reports the confusion matrix applied test data set with cut-off point of $0.49$. According to the table, the predictive model also performs well. Secondly, receiver operating characteristic (ROC) is used to measure the performance of the predictive mode as the cut-off point is varied. 
This curve is created by true positive rate and false positive rate. The diagonal is called random guess line since that if we guess randomly, $FPR$ and $TPR$ are $50\%$ to $50\%$ (Figure \ref{roc}). The ROC space is the distance between point on ROC curve and line. The best possible prediction yield the left corner point $(0,1)$ which is also named perfection classification. In this test, the area under ROC curve achieves $0.95$. 

\mytable{
	\begin{tabular}{c   c | c c }
		\cline{3-4}
		&  & \multicolumn{2}{ c| }{Predictive} \\ \cline{3-4}
		\multicolumn{1}{c }{} & & False & 
		\multicolumn{1}{ c|}{True}\\  \cline{1-4}
		\multicolumn{1}{ |c|}{\multirow{2}{*}{Actual}} & 0 & 29820 (TN) & 
		\multicolumn{1}{ c|}{5363 (FN)} \\
		\multicolumn{1}{ |c|}{} & 1 & 221 (FP) & 
		\multicolumn{1}{ c|}{2246 (TP)} \\
		\hline
	\end{tabular}
}{Confusion Matrix with Cut-off $0.49$ of Test Dataset} {CMT}

In summary, through rigorous validation process, it can be inferred that the logistic regression model is accurate on both train data set and test data set. The final model is displayed in Table \ref{logistic} and the cut-off point is set as $0.49$ in the Chapter \ref{Applications}.

\section{Interpretation of Logistic Regression Model}
\label{Interpret}

This section interprets the notations and the corresponding values in the logistic regression model in detail. Firstly, the model is in agreement with some intuitive network characteristics and quantifies the exact effect statistically.
In Table \ref{logistic}, $m$ denotes the total number of arcs in the network and is statistical significant in terms of the low p-value ($0.003$) and the high significant code (`***' refers to the most significance). The logistic parameter of $m$ ($\beta_{m}$) equals $-5.52E-03$ with the standard error of $1.86E-03$ and the corresponding odds ratio is $0.994$, which is obtained by taking the exponential of $\beta_{m}$. This is to say, holding every predictors constant except $m$, when the network has one more arc, the response variable decreases by $0.5 \%$. Intuitively, more arcs in the network results in more alternative paths exiting in the network. Therefore, the overall probability for an arc to be selected in the optimal solution is naturally negative correlated with the number of arcs. 
Both variable and fixed costs have a significantly  effect on the response variables as showed in Table \ref{logistic}. The $1$-unit increase of variable cost and fixed cost can drop the response variable by $8.4 \%$ and $0.01\%$, respectively. However, we can not directly infer that variable costs are more influential than fixed costs, since the odds ratio should be combined with the range of the predictors to evaluate the impact. As shown in Table \ref{statis}, the range of $c_{ij}$ is $10$ and the corresponding  of response variable has  maximum $84\%$ increase/decrease, while the $f_{ij}$ can vary between $20,000$ and $60,000$, hence the possible corresponding probability can increase/decrease by $400 \%$ at most. Consequently, fixed cost has more effect with above variety with this range. 

Secondly, two hidden findings can be concluded from the model. Although fixed and variable costs are statistical significant, the ratio between them is not so important because the p-value is $0.27$ and the odds ratio of $\gamma_{ij}$ is $1.000000128$, close to $1$. In addition, total supply requirements and density are involved in our candidate predictors, but they are not statistically significant in terms of Akaike information criterion. Therefore, the accuracy of the predictive model is not affected by total supply requests and the network density.

Thirdly, an arc is whether or not used in the FCNF is notably affected by the corresponding head and tail nodes because almost all the features of nodes are statistically significant. For example, the response variable increases by $779 \%$ if the type of head node changes from demand to supply. This could also be verified by the odds ratio of $\bar{r}_{i}$, which indicates that $1$-unit increase of the requests of head node leads to $162\%$ increase of response variable. Keeping the other predictors constant, if the indegree of the head node increase $1$-unit, there is $93 \%$ decrease of the response variable. The other relationships can be found in Table \ref{logistic}.

Finally, an arc has positive flow in the linear relaxed FCNF has a much higher probability to be used in the original FNCF. With the notable influence of $\bar{l}_{ij}$, the response variable is $3,1730 \%$ more likely to be used when the flow of arc increases from zero to full capacity. If an arc is open in the linearized FCNF, the likelihood of this arc has positive flow increases by $307\%$. This property is commonly employed to improve the branch and bound technique at either root node or branch nodes \citep{fischetti2003local,danna2005exploring, lodi2010mixed, zhang2014OSEA}. This paper provides statistical evidence to support these works but also indicates the non-negligible difference between the optimal solutions of the linear relaxation problem and the original FCNF. In order to illustrate the problem, we build a logistic regression model on the train dataset only with one predictor ( $\bar{l}_{ij}$) and the confusion matrix is showed in Table \ref{CML}. The total false cost of Table \ref{CML} is $0.21$, while the logistic regression model (Table \ref{logistic}) is only $0.11$, in which other predictors revise the misclassification of $\bar{l}_{ij}$. All these findings can be applied to analyze the networks characteristics, identify critical components of the network and provide preprocessing information to the FCNF problem.

\mytable{
	\begin{tabular}{c   c | c c }
		\cline{3-4}
		&  & \multicolumn{2}{ c| }{Predictive} \\ \cline{3-4}
		\multicolumn{1}{c }{} & & False & 
		\multicolumn{1}{ c|}{True}\\  \cline{1-4}
		\multicolumn{1}{ |c|}{\multirow{2}{*}{Actual}} & 0 & 6187 (TN) & 
		\multicolumn{1}{ c|}{395 (FN)} \\
		\multicolumn{1}{ |c|}{} & 1 &  2844 (FP) & 
		\multicolumn{1}{ c|}{3923 (TP)} \\
		\hline
	\end{tabular}
}{Confusion Matrix of logistic regression with only $l_{ij}$ on Train Dataset} {CML}

\chapter{Applications of Predictive Model}
\label{Applications}

\section{Important Index and Case Study}
\label{EIIC}

According to literature review in Section \ref{ICC}, to identify the critical components or develop components importance index (CII), a measure is required to evaluate the network performance. In this thesis, the objective value of the optimal solution in FCNF is employed to quantify the network service level. The objective value is not a single cost value, but results through solving the FCNF. The optimal solution considers many metrics that are widely used individually to evaluate the network, e.g. network flow, link capacity, delivery costs and network connectivity. Since FCNF is NP-hard and computational expensive to find the optimal solution, no existing literature uses it to measure the network performance. When the problem extends to identify critical groups of edges of different sizes over the entire network, it is a NP-complete combination problem, which is difficult to solve by itself. The predictive model provides a efficient, simple, inexpensive and practical way to compute the CII for each arc in terms of FCNF without solving any NP-complete or NP-hard problems, which extends the metrics of evaluating network performance family. 

The CII is defined as the probability that an arc is selected in the optimal solution of the FCNF for the reason that, intuitively, destruction or damaging of arcs in the optimal solutions causes a rerouting of the flow from optimal path to a non-optimal path with higher costs between supply and demand nodes.  
The higher probability indicates the arc is more important and vice verse. To formulate the measure, let $\pi_{ij}$ denote the likelihood that arc $(i,j) \in A$ is used in the optimal solution, the value of which is directly from the predictive model developed in Section \ref{AICselection}. Let $b_{ij}$ represent whether or not to remove arc $(i,j)\in A$ from the network, 
\begin{align*}
 b_{ij} = \begin{cases}
 0, \text{remove arc (i,j)}\\
 1, \text{otherwise}
 \end{cases}
 \forall (i,j) \in A.
\end{align*}

Furthermore, let $z=z(b_{ij})$ where $z(b_{ij})$ represents the relationship between total cost of the network and removal status of arc $(i,j)\in A$,
\begin{align*}
z(b_{ij}) =  \text{min} & \sum_{(i,j) \in A} (c_{ij}x_{ij} + f_{ij}y_{ij})   \\
\text{s.t.} & \sum_{(i,j) \in A} x_{ij} - \sum_{(j,i) \in A} x_{ji} = R_{i}&
\forall{i \in N}  \\
& 0 \leq x_{ij} \leq M_{ij}y_{ij}&
\forall{(i,j) \in A}  \\
& y_{ij} \in \lbrace 0,1 \rbrace&
\forall{(i,j) \in A} \\
& y_{ij} \leq b_{ij} &
\forall{(i,j) \in A} \\
& b_{ij} \in \{0,1\} &
\forall (i,j) \in A
\end{align*}

Finally,  we define the failure effect as the percentage increase of cost caused by the disruption of segments compared with the cost incurred when all arcs are present in the network ($b_{ij}=0, \forall(i,j)\in A$) represented by $z_0$. Let $\eta(b_{ij})$ denote the failure effect and the equation is expressed as,
\begin{align*}
\eta(b_{ij}) = \dfrac{z(b_{ij})-z_0}{z_0}\times 100 \% 
\end{align*}

In real-world, it is not practical to solve the large FCNF problem exactly and the value of $\eta_{ij}$ is unavailable due to the computational difficulty. According to our experiments, the solving time increase significantly when the number of nodes achieves $30$. However, the predictive model is simply linear calculation and can output the results immediately. 
The critical components measure approach is illustrated with a random directed network as depicted in Figure \ref{node10}. Node 1, 7 and 9 are supply nodes and node 0, 5 and 6 are demand nodes. The arrow on the line indicates the direction and each node-pair has two directed arcs, thus we total have $28$ arc variables. The value of $z_0$ for this instance is $316074$.

\myfigure{
	\centering{\includegraphics[width=0.8\textwidth]{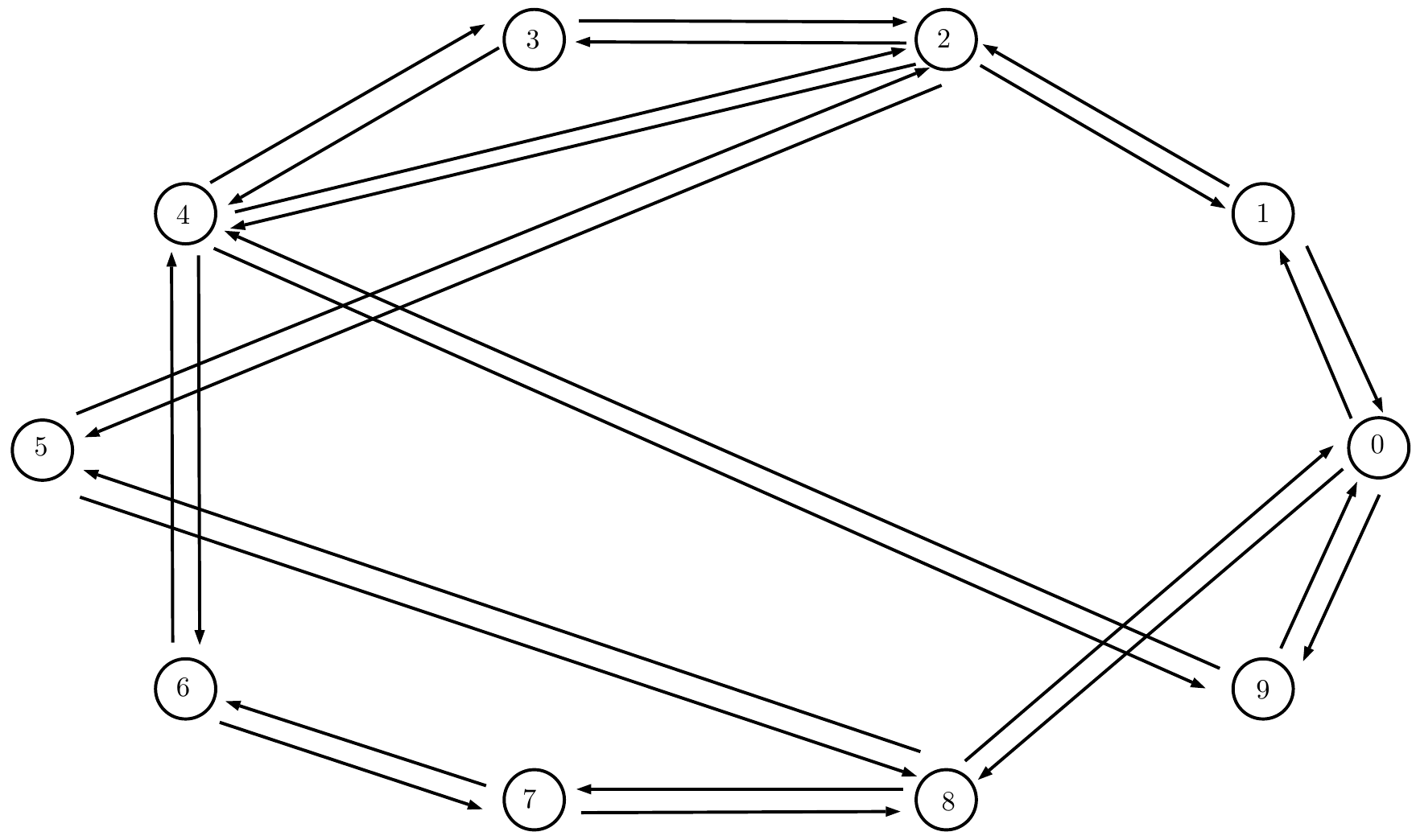}}
}{Small Directed Network with $10$ Nodes and $28$ Arcs}{node10}
 
Table \ref{PF} displays the variable cost, fixed cost, probability to be in optimal solution, failure effect for each arc comprising the network. Note, since this is a directed network, arc $(0,1) \in A$ is not same as arc $(1,0) \in A$. Based on this table, both the probability of arc $(7,6)$ and $(9,0)$ is $0.99$, and their failure effects are $10.22\%$ and $6.78\%$, respectively. Although they do not have the lowest variable cost or fixed cost, the endpoints of these two arcs are a pair of supply and demand nodes. The predictive model automatically considers all the features of this arc and give the probability that the likelihoods of these two arcs to be selected in the optimal solution are close to $1$.  Some arcs with low costs, e.g. $(6,7)$, $(0,9)$, $(8,0)$, $(5,8)$ and $(8,7)$, have high CII values as well. These arcs can be contrasted with arcs with very low CII values, which means they do not contribute to any failure effects. In addition, according to the table, the directed arcs connecting same pair of nodes do not share same failure effect and probability. For example, the value of probability and failure effect of arc $(9,4)$ are much higher than arc $(4,9)$. Consequently, it is non-trivial to consider the direction of arcs in the identification of critical components.
Finally, optimization approach to \citet{zio2012identifying, shen2012exact} can only find the optimal critical segments in the network, but the predictive model is capable to calculate CII for each bridge. Accordingly, decision maker can combine the CII with other indices, e.g. social and economic, to select critical components.

\mytable{
	\begin{tabular}{c  c c  c c}
		\hline
		Arc$(i,j)$ & $c_{ij}$ & $f_{ij}$ & $\pi_{ij}$ & $\eta_{ij}$ \\
		\hline
		(7,6)	&	7.20	&	41011.33	&	0.99	&	10.22	\%	\\
(9,0)	&	6.51	&	44846.94	&	0.99	&	6.78	\%	\\
(9,4)	&	5.95	&	22028.81	&	0.95	&	7.81	\%	\\
(1,0)	&	9.69	&	42158.06	&	0.95	&	7.81	\%	\\
(6,7)	&	1.93	&	24561.20	&	0.82	&	10.22	\%	\\
(1,2)	&	7.17	&	44364.01	&	0.73	&	0.37	\%	\\
(0,9)	&	3.10	&	28765.87	&	0.66	&	6.78	\%	\\
(8,0)	&	7.54	&	26588.18	&	0.49	&	4.54	\%	\\
(2,5)	&	8.90	&	52552.74	&	0.45	&	0.37	\%	\\
(5,8)	&	3.90	&	28365.83	&	0.35	&	7.81	\%	\\
(7,3)	&	1.47	&	43307.38	&	0.23	&	7.81	\%	\\
(8,7)	&	0.13	&	33349.63	&	0.17	&	7.81	\%	\\
(4,6)	&	9.09	&	53894.90	&	0.11	&	4.54	\%	\\
(3,7)	&	4.93	&	40543.16	&	0.09	&	0.37	\%	\\
(7,8)	&	5.68	&	49949.96	&	0.09	&	0.00	\%	\\
(0,1)	&	1.27	&	56093.72	&	0.04	&	0.00	\%	\\
(2,3)	&	5.90	&	31594.21	&	0.03	&	0.00	\%	\\
(5,2)	&	2.53	&	46060.27	&	0.03	&	0.00	\%	\\
(4,9)	&	1.21	&	48247.28	&	0.03	&	0.00	\%	\\
(8,5)	&	4.01	&	57456.67	&	0.02	&	0.00	\%	\\
(0,8)	&	8.61	&	47039.03	&	0.02	&	0.00	\%	\\
(3,2)	&	3.43	&	41127.17	&	0.01	&	0.00	\%	\\
(2,1)	&	5.07	&	50998.22	&	0.01	&	0.00	\%	\\
(6,4)	&	7.67	&	48589.75	&	0.01	&	0.00	\%	\\
(4,2)	&	7.62	&	39342.56	&	0.01	&	0.00	\%	\\
(4,3)	&	4.77	&	46389.37	&	0.00	&	0.00	\%	\\
(3,4)	&	6.18	&	48528.36	&	0.00	&	0.00	\%	\\
(2,4)	&	3.63	&	54166.87	&	0.00	&	0.00	\%	\\
		\hline		
	\end{tabular}
}{Probability and Failure Effect for Each Arc in Network} {PF}

In order to evaluate the quality of critical identification method, the experiment design includes tests on a variety of network densities, each which has characteristics corresponding to difficult FCNF problem instances (e.g. high fixed to variable cost ratio). The tests consists of $100$ networks with $20$ nodes, for which we can solve optimally in a reasonable time horizon. Specifically, we create a connected network with the number of arcs randomly selected between $38$ to $380$. In this experiment, we perform comparison of failure effect by blocking two arcs with highest CII and two arcs with lowest CII. 
The distribution of failure effect by removing top two critical arcs is displayed in Figure \ref{DFE}. Among all the $100$ instances, the blockage of the two critical arcs leads to average $9.85\%$ increase of network cost and in some cases, the failure effect achieves over $30\%$. However, the failure effects of two non-critical arcs are average to $0.21\%$ and in the $94$ instances among tests, the failure effects are $0\%$. Consequently, it is reasonable to conclude that the arcs with higher CII values can lead to more failure effects and should be protected as critical segments.

\myfigure{
	\centering
	\includegraphics[width=0.8\textwidth]{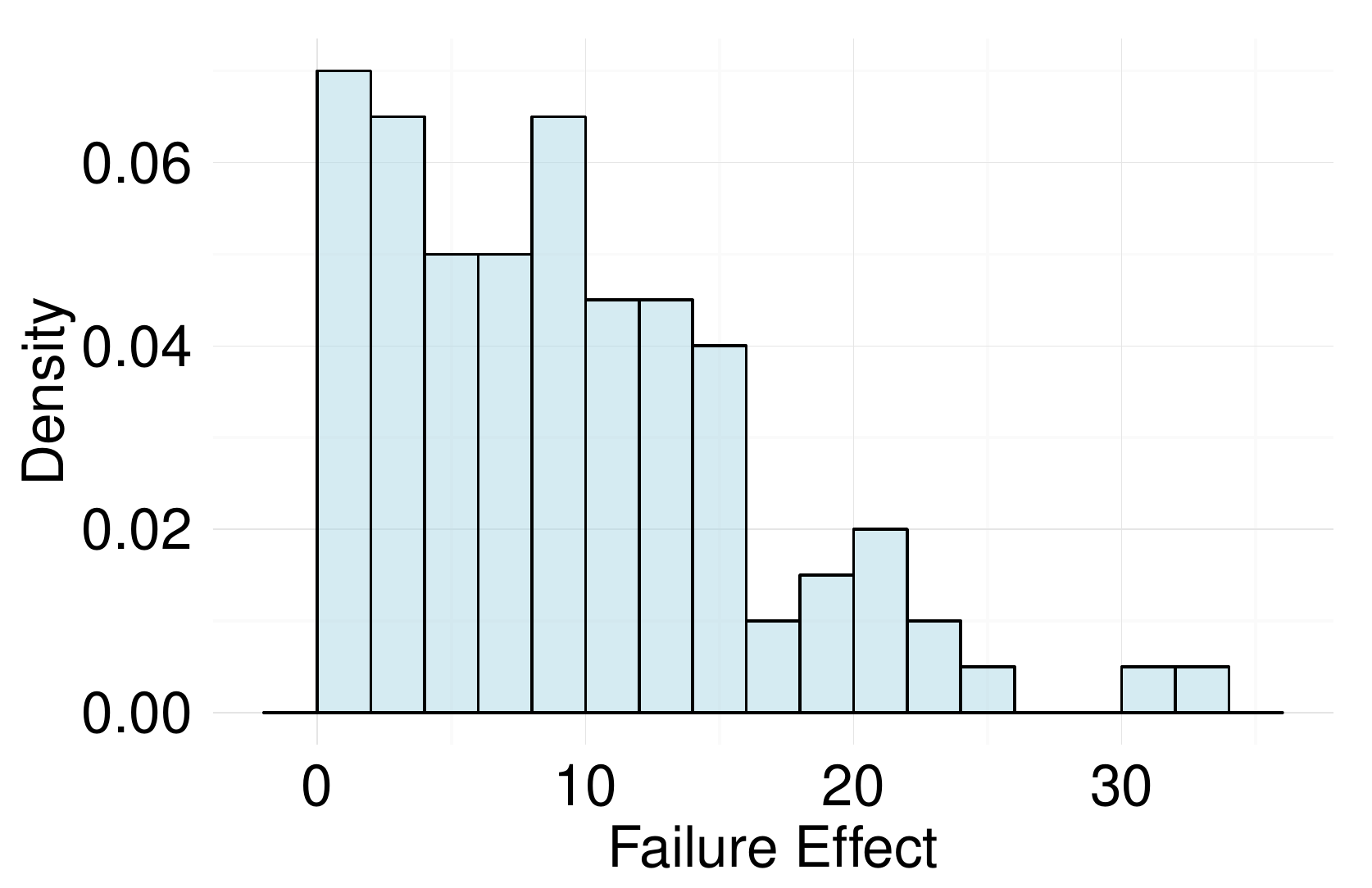}
}{Distribution of Failure Effect by Removing Top Two Critical Arcs}{DFE}

\section{Regression Based Relaxation}
\label{Regression Based Relaxation}

\subsection{RBR Formulation}
\label{RBRformulation}

In this section, a novel approximate solution approach to the FCNF based on the predictive model is discussed, which is named regression-based relaxation (RBR).  Let $p_{ij}$ denote the probability that arc $(i,j) \in A$ is open in the FCNF problem, which is calculated by the predictive model.
Let $c^{'}_{ij} = -\ln p_{ij}$ where $0 < p_{ij} \le 1 (\forall (i,j) \in A)$ and  $z'_\text{RBR}$ denote the objective value of the RBR FCNF, which is defined as
\begin{align*}
z'_\text{RBR} &= \sum_{(i,j) \in A} c^{'}_{ij} x_{ij} \\
&= \sum_{(i,j) \in A} -x_{ij} \ln p_{ij}  \\
&= -\ln \prod_{(i,j) \in A} p_{ij}^{x_{ij}} 
\end{align*}
\noindent The RBR problem formulation is: $\min z'_\text{RBR}$ subject to \eqref{equilibrium} -- \eqref{binary}. In a simple case with single supply node $s$, single demand node $t$, and total network supply equal to 1 in a feasible problem, the RBR solution will be a \emph{most probable feasible path} from $s$ to $t$.  That is, if $P_{st}$ denotes a most probable feasible path from $s$ to $t$, $$\min z'_\text{RBR} = - \ln \prod_{(i,j) \in P_{st}} p_{ij}$$ and $x_{ij} = 1 \iff x_{ij} \in P_{st}$.  In general for feasible FNCF problems with $\mathbf{x} \ge 0$, the solution to RBR identifies a set of arcs which form feasible likely paths from possibly many supply nodes to many demand nodes.

Here, the RBR problem is linear and can be solved immediately. The optimal solution of RBR is easily transformed to a feasible solution of the original FCNF. Let $x_{ij}^*$ denote the value of optimal flows on arc $(i,j) \in A$ in the RBR problem, and the corresponding decision variables, $y_{ij}^*$ , are assigned as,
\begin{align}
y^*_{ij} = \begin{cases}
1, x^*_{ij}>0\\
0, \text{otherwise}
\end{cases} \forall{(i,j) \in A}. \label{update}
\end{align}
The updated solution satisfies all the constraints in the FCNF. Let the objective value of the FCNF found using RBR be denoted by $z_{RBR}$ and calculated by,
\begin{align*}
z_{RBR} = \sum_{(i,j) \in A} (c_{ij}x^*_{ij} + f_{ij}y^*_{ij}).  
\end{align*}

\subsection{RBR Computation Results}
\label{RBRresults}

In order to evaluate the solution quality and efficacy of the RBR approximate solution approach to the FCNF, we randomly generate 626 FCNF instances as the test bed, which includes a large variety of network sizes and all the instances have characteristics with respect to difficult FCNF (high fixed and variable cost ratio). The tests are classified into three levels by the number of nodes, 224 easy ($10 - 300$ nodes), 236 medium ($350-650$ nodes), and 166 hard ($700-1000$ nodes) problems. The number of arcs, $m$, is randomly selected. The percentage of supply, demand and transshipment nodes are randomly chosen with respectively approximate probabilities, $0.2$, $0.2$, and $0.6$. The probabilities are approximate because adjustments are necessary to ensure the instance is feasible. The variable costs and fixed costs follow $\text{U}(0,10)$ and $\text{U}(20000, 60000)$, respectively. The total number of supplies is randomly assigned on $\text{U}(1000, 2000)$. 

Let $\rho_{s}$ and $\rho_{d}$ denote the percentage of supply and demand nodes, respectively. Table \ref{test} reports the statistics of the network characteristics of the test bed (Min., 1st Qu., Median, Mean, 3rd Qu. and Max.). The smallest instance is composed of 10 nodes and 20 directed arcs, which is close to the problem size of train dataset, whereas the largest problem contains $1,000$ nodes and $82,010$ arcs, which is extremely lager than the network in train and test dataset. Furthermore, the first quarter of number of nodes is $250$ which is also much larger than the largest problem in train and test dataset. The average supply quantity ranges from $333$ to $1,971$. The ratio of fixed costs to variable costs averages to $8,008$. The test bed includes FCNF instances with $10 \%$ to $50 \%$ of nodes as supply nodes, and $8\%$ to $40\%$ as demand nodes.

The experiments employs linear programming (LP) relaxation and state-of-the-art exact technique as the benchmarks. 
The objective value of the FCNF found using LP relaxation is denoted by $z_{\text{LP}}$. We select state-of-the-art Gurobi software as the exact optimization technique, which is a commercial optimization (linear, integer and mixed integer programming) software. By default setting, Gurobi uses 14 different MIP heuristics, 16 cutting plane strategies, and several presolve techniques \citep{optimization2012inc}. The best objective value found using Gruobi 5.6.3 is denoted as $z_{\text{GRB}}$. The experiments are performed on a Windows 7 64bit machine with Intel Xeon CPU E5-1620 and 8 GB RAM. The time limit for all three techniques is 60 seconds and the actual running time for RBR, LP and Gurobi are recoded as $t_{\text{RBR}}, t_{\text{LP}}$ and $t_{\text{GRB}}$, respectively. 

\mytable{
	\begin{tabular}{c  | c c  c c c c}
		\hline
		Paramters	&	Min.	&	1st Qu.	&	Median	&	Mean	&	3rd Qu.	&	Max.	\\
		\hline
		$n$	&	10	&	250	&	400 &   446.7 &  700 & 1000	\\
		$m$	&	20  &   9497  & 26270  & 29520 &  46480  & 82010	\\
		$\rho$	&	0.03	&	0.24	&	0.58	&	0.55	&	0.89	& 1.00	\\
		$\bar{S}$	&	333	&	858.2	&	947	&	968.4	&	1039	&	1971	\\
		$\gamma$	&	7110	&	7986	&	8000	&	8008	&	8010	&	9809	\\
		$\rho_{s}$ & 0.1000 &  0.3000 & 0.3178 & 0.3173 & 0.3329  & 0.5000 \\
		$\rho_{d}$ & 0.0800 & 0.1887 & 0.2000 & 0.2033 & 0.2161 & 0.4000 \\
		\hline		
	\end{tabular}
}{Statistics Information of the Test Bed} {test}

In this section, we report the analysis of solution quality and efficiency of RBR, LP and Grurobi. Let $z_{\text{gap}}^{x}$ denote the percentage gap between $z_{\text{RBR}}$ and $z_{x}$,
\begin{align}
z_{\text{gap}}^{x} = \dfrac{z_{\text{RBR}}-z_{x}}{\vert z_{x}\vert} \times 100 \%.  \label{gap}
\end{align}
where $z_{x}$ is one of $z_{\text{LP}}$ or $z_{\text{GRB}}$. Let $t_{\text{diff}}^{x}$ denote how many times RBR is faster/slower than $x \in \{\text{LP}, \text{GRB}\}$,
\begin{align}
t_{\text{diff}}^{x} = \dfrac{t_{x}}{ t_{\text{RBR}}},  \label{diff}
\end{align}
here, if $t_{\text{diff}}^{\text{LP}}$ equals to $2$, then solving the RBR problem is twice faster than solving the LP problem; reversely, if $t_{\text{diff}}^{\text{LP}}$ equals to $0.5$, then $t_{\text{RBR}}$ is twice longer than $t_{\text{LP}}$. 

The performance is measured with respect to $z_{\text{gap}}^{x}$ and $t_{\text{diff}}^{x}$ together. Based on the formulation (\ref{gap}), if RBR outperforms LP (Gurobi) in terms of objective value, the value of $z_{\text{gap}}^{LP}$ ($z_{\text{gap}}^{GRB}$) is negative; if RBR outperforms LP (Gurobi) with respect to CPU time, the value of $t_{\text{diff}}^{LP}$ ($t_{\text{diff}}^{GRB}$) is larger than $1$. Let binary variable $\theta_{x}$ denote whether RBR outperforms $x \in \{\text{LP}, \text{GRB}\}$, 
\begin{align*}
\theta_{x} = [z_{\text{gap}}^{x}<0][t_{\text{diff}}^{x}>1]
\end{align*}
where, $[X]$ returns $1$ if $ X $ is true; otherwise, returns $0$. In this investigation, we claim RBR outperforms LP (Gurobi) only when $\theta_{\text{LP}}=1$ ($\theta_{\text{GRB}}=1$), respectively. Table \ref{statistics} reports the experimental results by techniques.


\mytable{
	\begin{tabular}{c c c c c c c}
		\hline
		& & \multicolumn{3}{c}{Levels} & \\ \cline{3-5}
		& & 	Easy	 & 	Medium	 & 	Hard	 & 	Overall	\\
		\hline
		\multirow{7}{*}{RBRvsLP} & $\theta_{\text{LP}} = 1 ( \%)$	 & 	91.96	 & 	100	 & 	100	 & 	97.12	\\
		\cline{2-6}
		& min $z_{\text{gap}}^{\text{LP}} ( \%)$ 	 & 	-43.39	 & 	-45.05	 & 	-46.94	 & 	-46.94	\\ [1.5ex]
		& mean $z_{\text{gap}}^{\text{LP}} ( \%)$ 	 & 	-25.23	 & 	-35.57	 & 	-36.18	 & 	-32.03	\\ [1.5ex]
		& max $z_{\text{gap}}^{\text{LP}} ( \%)$ 	 & 	0.00 	 & -11.96 	 & -7.31	 & 	 0.00	\\ [1.5ex]
		\cline{2-6}
		& min $t_{\text{diff}}^{\text{LP}}$ 	 & 	1.26	 &  3.17	 & 	2.93	 & 	1.26	\\ [1.5ex]
		& mean $t_{\text{diff}}^{\text{LP}}$ 	 & 	24.65	 & 	41.48	 & 	26.55	 & 	31.5	\\ [1.5ex]
		& max $t_{\text{diff}}^{\text{LP}}$ 	 & 	147	 & 186.2 & 	116.9	 & 	186.2	\\ [1.5ex]
		\hline
		\multirow{7}{*}{RBRvsGRB} & $\theta_{\text{GRB}} = 1 ( \%)$	 & 	5.28	 & 	46.41	 & 	98.05	 & 51.69	\\
		\cline{2-6}
		& min $z_{\text{gap}}^{\text{GRB}} ( \%)$ 	 & 	-20.60	 & 	-41.17	 & 	-44.93  & 	-44.93	\\ [1.5ex]
		& mean $z_{\text{gap}}^{\text{GRB}}( \%)$ 	 & 	14.29	 & 	 -1.25	 & 	-29.84	 & 	-6.16	\\ [1.5ex]
		& max $z_{\text{gap}}^{\text{GRB}}( \%)$ 	 & 	51.20	 & 32.68 & 	 14.62 	 & 	51.20	\\  [1.5ex]
		\cline{2-6}
		& min $t_{\text{diff}}^{\text{GRB}}$ 	 & 	3.31 	 & 	87.41	 & 	 11.28	 & 	3.31	\\ [1.5ex]
		& mean $t_{\text{diff}}^{\text{GRB}}$ 	 &  5756	 & 	591	 & 	177	 & 	2052	\\[1.5ex]
		& max $t_{\text{diff}}^{\text{GRB}}$ 	 & 	82430	 & 5752	 & 	2791	 & 	82430	\\ [1.5ex]
		\hline
	\end{tabular}
}{Objective Gap and Time Difference Statistics} {statistics}

Overall, RBR outperforms LP in $97.12\%$ of the test bed and is never outperformed by LP (the maximum value of  $z_{\text{gap}}^{\text{LP}}$ is $0.00\%$). Figure \ref{LPgap} shows the distribution of objective gap between RBR and LP by difficult levels.  RBR finds objective values average $32.03 \%$ lower than LP relaxation solution approach. The percentage of FCNF instances in which RBR outperforms LP in both solution quality and efficiency achieves $100 \%$ for the medium and hard problems. For some specified instances, RBR performs quite well and the highest improvement of RBR to LP achieves $46.94 \%$. In addition, running time of solving a RBR problem is overall $31.5$ times faster than solving a LP relaxation problem. Based on the minimum value of $t_{\text{diff}}^{\text{LP}}$ ($1.26$), there does not exist a problem instance in which RBR uses longer time than LP. Figure \ref{LPtime} plots the running time as a function of the number of variables, from which it is observed that the running time of RBR is not affected apparently by the size of the problem instances (average $t_{\text{RBR}}=0.33$ seconds), albeit the running time of LP appears exponential relationship with the number of variables (average $t_{\text{LP}}=12.75$ seconds). 
Specifically, LP performs equally with RBR only when the number of nodes is less than $10$. Therefore, RBR is guaranteed to find a better solution in a shorter time period than LP regardless of the scale and complexity of FNCF problem.

\myfigure{
	\centering
	\includegraphics[width=1\textwidth]{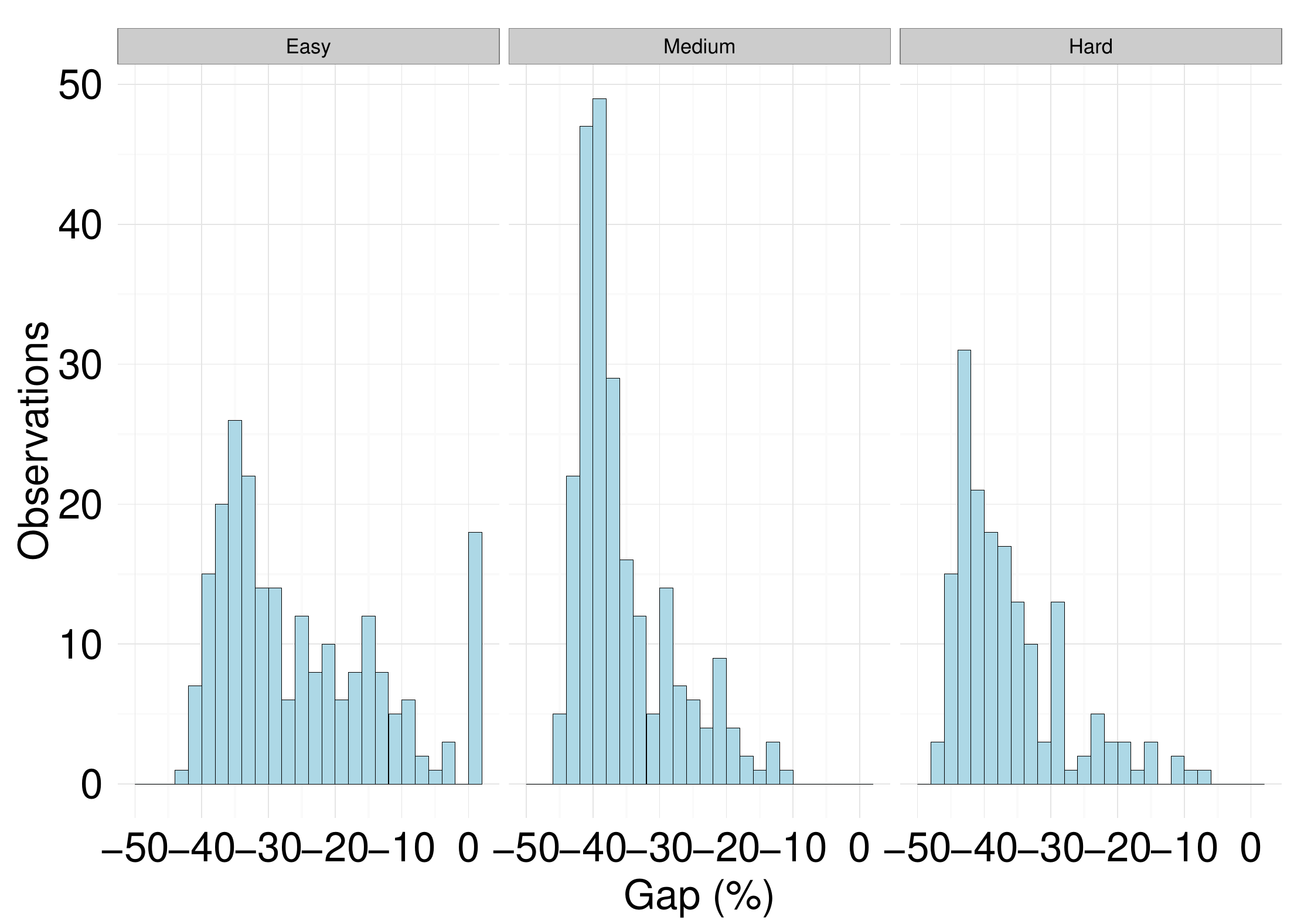}
}{Distribution of Gap between $z_{RBR}$ and $z_{\text{LP}}$ by Level }{LPgap}

\myfigure{
	\centering
	\includegraphics[width=1\textwidth]{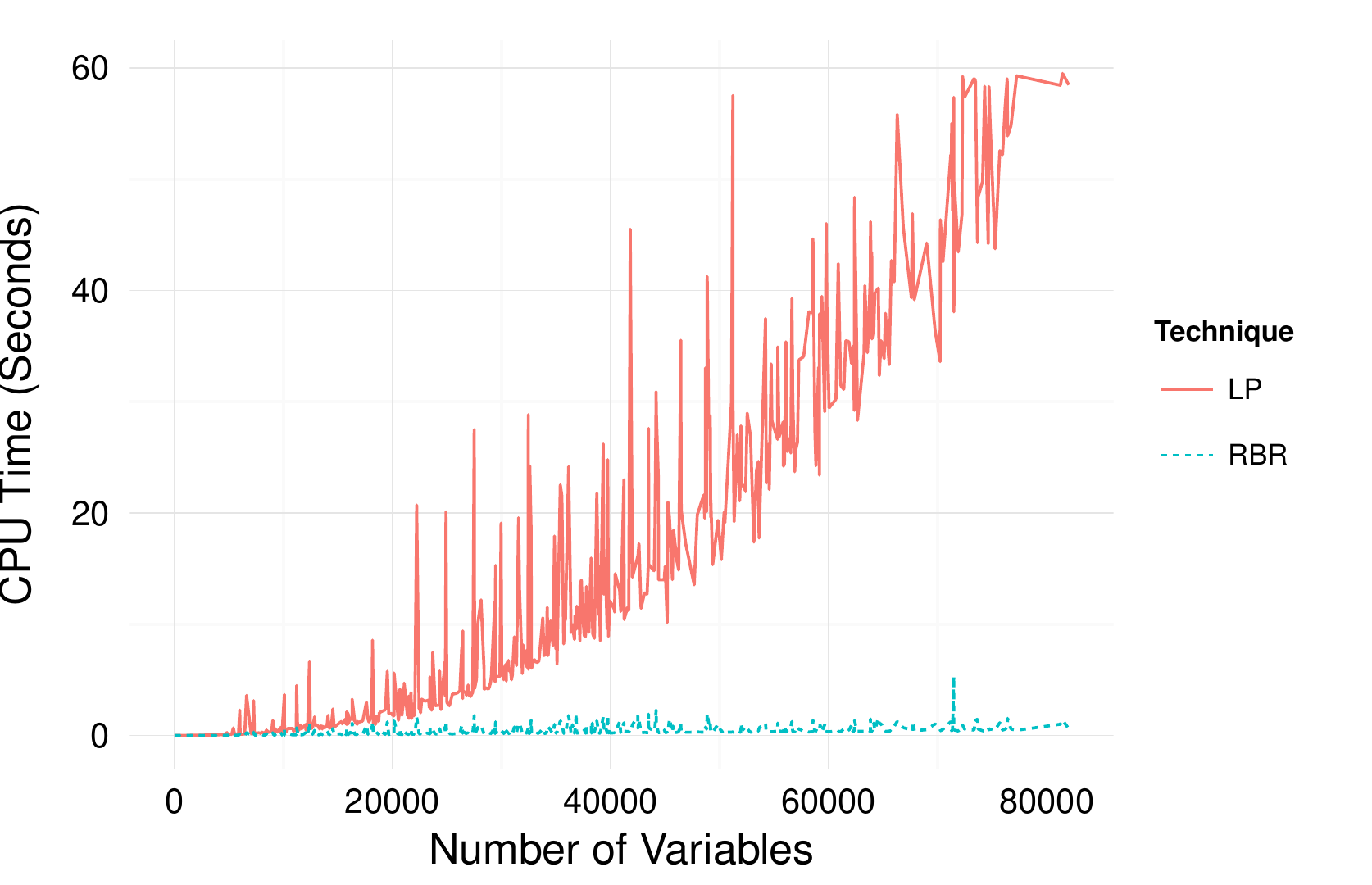}
}{ CPU Time of RBR and LP over Number of Variables}{LPtime}

State-of-the-art exact solver outperforms RBR with respect to solution quality in $95 \%$ of the test instances, but solving RBR problem is overall $5,756$ times faster than GRB with 60 seconds limit. In detail, GRB achieves the time limit ($60$ seconds) when the number of variables is $220$, in which RBR only runs around $0.001$ seconds. The solution produced by RBR begins showing comparable to Gurobi when the difficult level is medium. Figure \ref{GRBgap} shows the gap distribution between $z_{RBR}$ and $z_{GRB}$ by difficult level. For hard FCNF problems, RBR outperforms Gurobi in $98\%$ of the hard instances and is only outperformed by Gurobi in $5$ over $248$ hard cases. The average improvement for hard instances is around $30 \%$ with much less running time. Since the running time of Gurobi is always extended to $60$ seconds while average RBR running time is only $0.33$ seconds, the investment in involving RBR as a pre-prosessing technique to the FCNF is efficient. Specifically, RBR could could find a better upper bound in the solving process and then reduce total running time of optimization solver.

\myfigure{
	\centering
	\includegraphics[width=1\textwidth]{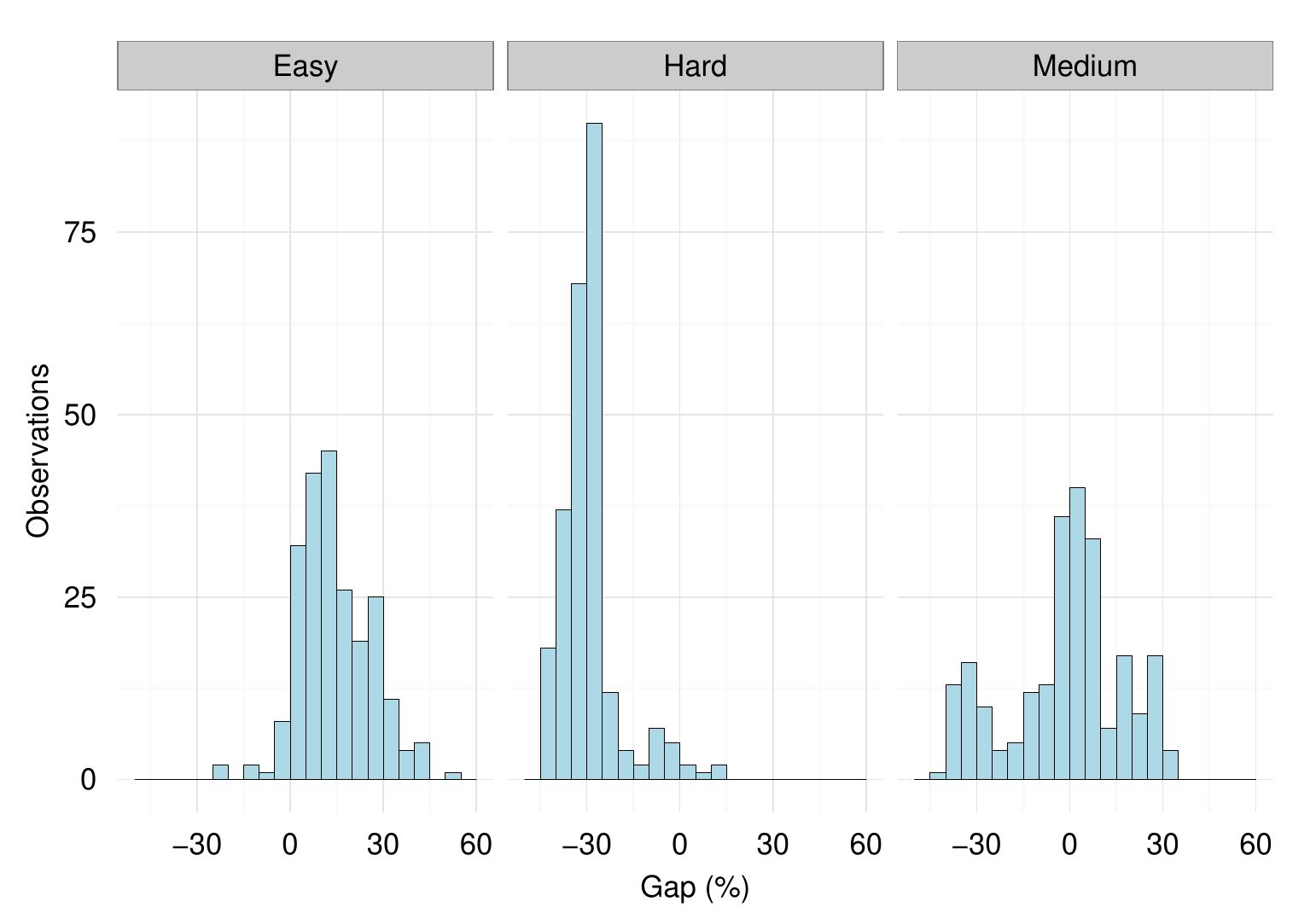}
}{Distribution of Gap between $z_{RBR}$ and $z_{\text{GRB}}$ by Level }{GRBgap}

\chapter{Conclusion}
\label{conclusion}

The fixed charge network flow problem has many real-world applications and can be transformed to various problem types. Because it is a classically NP-hard problem, many approximate algorithms have been developed to find a near-optimal solution in a reasonable time period. This research provides a novel and original framework to analyze the optimal flow of the FCNF. In this work, we address three questions in terms of the analysis, prediction and application.

Firstly, by analyzing the possible features that may affect the optimal flow, we follow three levels from the entire network to an arc and to a single node to ensure we extract all the useful features. There are $33$ independent features and only $26$ of them are employed as predictors for the final predictive model  with respect to Akaike information criterion. The validation process includes most common diagnostic techniques, e.g. confusion matrix, k-fold cross validation, ROC curve. The results indicate that the predictive ability is highly accurate on both train dataset and test dataset. According to the model, we can conclude that: (1) as the increase of network size, the average probability for each arc decreases; (2) the characteristics of arcs (e.g. variable and fixed costs) and nodes (e.g. node type and degree) have a statistical significant effect on the likelihood; (3) the logistic regression model is in agreement with that the linear relaxation and original problem share a significant proportion of optimal arcs. However, our study also finds that the linear relaxed FCNF loses considerable number of optimal arcs in the original problems. The rest predictors play an important role on improving the predictive capability. It is observed that the types of nodes of the arc and endpoints adjacent to the nodes provide a vital contribution in reducing the predictive errors.

Secondly, this predictive model can be used directly to identify critical arcs in the network. The component importance index (CII) is defined as the likelihood that the arc is selected in the optimal solution and the network performance is evaluated by the objective value of the FCNF. In our rigorous tests, the failure effects cased by high-CII arcs are statically significant than cased by low-CII arcs. The FNCF has never been used to measure the network performance due to the computational difficulty, not to mention quantifying the failure effect for each arc. The values of CII in our experiments are consistent with the failure effects and rank all the arcs, which supports the decision make on critical components identification or priorization.

Finally, the other application of the predictive mode is presented in the work, which we name regression-based relaxation (RBR). The RBR solution approach replaces the objective function with the natural logarithm of the product of probabilities calculated by predictive model. Then, the optimal solution of RBR problem is the most probable feasible path of the original FCNF problem. The solution quality of RBR is compared with linear programming relaxation solution approach, and state-of-the-art optimization software. The rigorous experiments show that RBR outperforms LP in $97.12\%$ of the test bed and produced objective values average $32.03\%$ lower than LP on average. Overall, solving a RBR problem is $31.5\%$ times faster then solving a LP problem. The empirical results indicates RBR stably outperforms Gurobi in the large-scale and high-complexity FNCF problems.


In summary, it is the first time to combine statistical learning technique and network flow problem. There are several vital network characteristics affecting the optimal flow regardless of the density and total requirements of the problem. The predictive model provides a novel approach to analyze and predict the optimal flow, which can be used independently on the identification of the critical components or approximate solution approach to the FCNF. 

The regression-based relaxation is one of possible applications and shows great advantage over the linear programming and standard exact technique. In the future work we will explore more applications that most benefit from this predictive model.

\chapter{Future Research}
\label{future}

Solution approach to discrete optimization has been popular for many years, but still a evolving research area. This research is the first to apply machine learning algorithms to study the optimal flows of the FCNF instead of transitional approach. We exploited the special structures of the FCNF problem and presented two possible applications. As part of our research, we identify the following important research topics that we plan to explore. 

1. \textbf{Network Science}. Currently, many research topics have transfered from single object to a group of homogeneous or inhomogeneous objects, e.g., transportation network, social network and community networks, etc. Network science, as a maturing field, offers a unique perspective to tackle complex problems, impenetrable to linear-proportional thinking.  The framework proposed in this work is for a specified optimization problem, but the concept or procedure can be applied to all types of networks with modification of features. In the future work, more types of network problems will be explored by this procedure and reversely, these networks can be employed to validate the methodology.   

2. \textbf{Solution Approach to difficult optimization problem}. Due to the success of regression-based relaxation, we plan to develop strategy combing the RBR and Branch-and-bound technique to provide a high efficiency exact solution approach to the FCNF problem. Furthermore, many difficult optimization problems has their own characteristics, such like Knapsack problem, traveling salesman problem, and job shop scheduling, etc, thus, we can study the parameters, decision variables and constraints of these problems and develop predictive model based on the known optimal solution. 

3. \textbf{Scala Machine Learning}. As the datasets growing exponentially, many machine learning techniques are not practical to apply because of the limitation of computational power. Scala machine learning is a relative new area in artificial intelligent area. According to the experimental results in this work, the predictive model can be trained on a small data set of easy-solve instances, and keep its highly accuracy for larger and more complex problems. This conclusion is same as Scala Machine Learning and we plan to explore methods to apply machine learning algorithms to big data more reliably and efficiently.

\newpage

\addcontentsline{toc}{chapter}{References}
\bibliographystyle{plainnat}
\bibliography{ref2}
\renewcommand{\refname}{References} 
\end{document}